\documentclass[twocolumn,superscriptaddress,nofootinbib]{revtex4-2}
\usepackage[utf8]{inputenc}
\usepackage{xcolor}
\usepackage{braket}  
\usepackage{graphics,graphicx,epsfig}
\usepackage{amssymb,amsfonts,amsmath}
\usepackage{ifthen}
\usepackage[utf8]{inputenc}
\usepackage[pdftex,hidelinks]{hyperref}
\usepackage{hyperref}
\graphicspath{{Figures/}}		

\usepackage{dsfont}
\newcommand{\EQ}{\begin{equation}}
\newcommand{\EE}{\end{equation}}
\newcommand{\EQA}{\begin{eqnarray}}
\newcommand{\EEA}{\end{eqnarray}}

\newcommand{\eff}{{\text{eff}}}

\newcommand{\cov}{\mathbf{cov}}
\newcommand{\rand}{{\text{rand}}}

\newcommand{\Q}{{\mathcal{Q}}}
\newcommand{\A}{{\mathcal{A}}}
\newcommand{\M}{{\mathcal{M}}}
\newcommand{\R}{{\mathcal{R}}}
\newcommand{\E}{\mathbb{E}}

\newcommand{\dd}{{\text d}}

\newcommand{\var}{\mathbf{var}}
\newcommand{\KL}{\mathrm{KL}}

\usepackage{hyperref}
\newcommand{\av}[1]{\left\langle{#1}\right\rangle}

\usepackage[normalem]{ulem}

\begin{document}

\title{Risk-utility tradeoff shapes memory strategies for evolving patterns}
\author{Oskar H Schnaack}
\address{Max Planck Institute for Dynamics and Self-organization, Am Fa\ss berg 17, 37077 G\"ottingen, Germany}
\address{Department of Physics, University of Washington, 3910 15th Ave Northeast, Seattle, WA 98195, USA}
\author{Luca Peliti}
\address{Santa Marinella Research Institute, 00058 Santa Marinella, Italy}
\author{Armita Nourmohammad}
\email{Correspondence should be addressed to Armita Nourmohammad: armita@uw.edu}
\address{Max Planck Institute for Dynamics and Self-organization, Am Fa\ss berg 17, 37077 G\"ottingen, Germany}
\address{Department of Physics, University of Washington, 3910 15th Ave Northeast, Seattle, WA 98195, USA}
\address{Fred Hutchinson Cancer Research Center, 1100 Fairview ave N, Seattle, WA 98109, USA}

\date{\today} 
\begin{abstract}
\noindent 
{Keeping a memory of evolving stimuli is ubiquitous in biology, an example of which is immune memory for evolving pathogens. However, learning and memory storage for dynamic patterns still pose challenges in machine learning. Here, we introduce an analytical energy-based framework to address this problem. By accounting for the tradeoff between utility in keeping a high-affinity memory and the risk in forgetting some of the diverse stimuli, we show that a moderate tolerance for risk enables a repertoire to robustly classify evolving patterns, without much fine-tuning. Our approach offers a general guideline for learning and memory storage in systems interacting with diverse and evolving stimuli.  } 
\end{abstract}
\maketitle


\section{Introduction}

Biological systems, ranging from the brain to the immune system, store memory of molecular interactions to efficiently recognize and respond to stimuli. Memory encoding in biological networks has also inspired a growing host of algorithms for learning and memory storage in image and pattern recognition by artificial neural networks~\cite{Goodfellow-et-al-2016,soltoggio_born_2018,Mehta:2019pr}. A critical step in these algorithms is to find regularities in data to associate related patterns with each other. As such, these learning algorithms often assume that the set of training data comes from a stationary distribution that represents the regularities necessary for pattern recognition in data well.  

Memory recognition, however, is not limited to static patterns and can be desirable when classifying evolving stimuli that drive the system out of equilibrium. One such example is the adaptive immune system in which memory  can effectively recognize evolved variants of previously encountered pathogens~\cite{Janeway:2001te,Barrangou:2014ht,AltanBonnet:2020hk,Bradde:2020kb,Schnaack:2020vb}. 
In a recent work, we have demonstrated that distributed learning strategies, which are desirable for pattern recognition in the stationary setup, can fail to reliably learn and classify dynamically evolving patterns~\cite{Schnaack:Hop1}. 
Specifically, we showed that to follow evolving patterns, an energy-based Hopfield-like neural network~\cite{Hopfield:1982fq} should use a higher learning rate, which in turn, can distort the energy landscape associated with the stored memory attractors, leading to pattern misclassification. To remedy this problem,  we proposed compartmentalized networks  as the optimal solutions to memory storage for evolving patterns~\cite{Schnaack:Hop1}. 

Irrespective of the network structure,  an increase in learning rate is necessary for a network to follow, recognize, and store effective memory of evolving patterns~\cite{Schnaack:Hop1}.  Increasing learning rate leads to a risky strategy, as the memory repertoire begins to reflect only the most recently encountered pattern while effectively destroying the memory of the prior encounters. Here, we present an analytical approach to explore how the tradeoff between utility and risk can determine learning strategies of a repertoire in keeping a memory of evolving patterns. We show that a moderate risk tolerance enables a repertoire to store an effective and robust memory. Our approach puts forward a guideline for optimal learning and memory storage for systems interacting with multiple evolving pattern classes without much fine-tuning.
\section{Model}

To probe memory strategies, we define a  repertoire $\M$ that can store stimuli (patterns) $\psi$ and later utilize them to recognize newly presented stimuli. We  consider  the space of possible  binary patterns  of length {$L$}, $\{\psi^\alpha \}$ with $\alpha \in {1,\dots,2^L}$ enumerating all unique patterns, with entries $\psi^\alpha_i = \pm 1$, for $i \in {1, \dots,L}$. The {memory repertoire $\M$} associates normalized weights $m^\alpha$ ($\sum m^\alpha = 1$)  to all patterns it encounters. The non-zero weights reflect the relative importance of the stored memory from different stimuli, and patterns that are not encountered  are associated with a zero weight. After an encounter with a given stimulus $\psi^\beta$ at time $t$, all weights are updated according to a Hebbian learning rule $m^\alpha(t) =(1-\lambda) m^\alpha(t-1) + \delta_{\alpha,\beta} \lambda $, where $\lambda$ is the learning rate~\cite{Hebb:1949vs}. Thus,  the weight $m^\alpha(t+\tau)$ associated with pattern $\alpha$, which was previously encountered at time $t$, decays  in an approximate exponential form over time,
\begin{align}
\label{eq:decay_time}
m^\alpha(t+\tau)= \lambda (1-\lambda)^\tau \approx\lambda e^{-\lambda \tau}.
\end{align}
\begin{figure*}[t!]
\centering
\includegraphics[width= 0.31\textwidth]{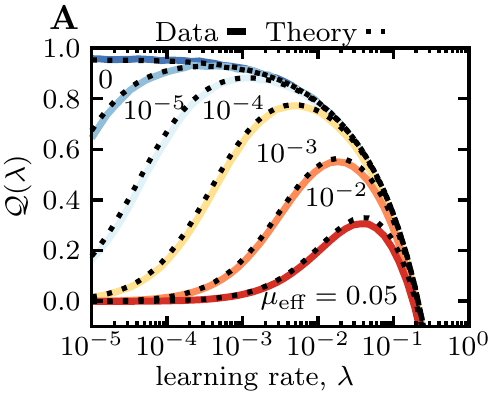}
\includegraphics[width= 0.33\textwidth]{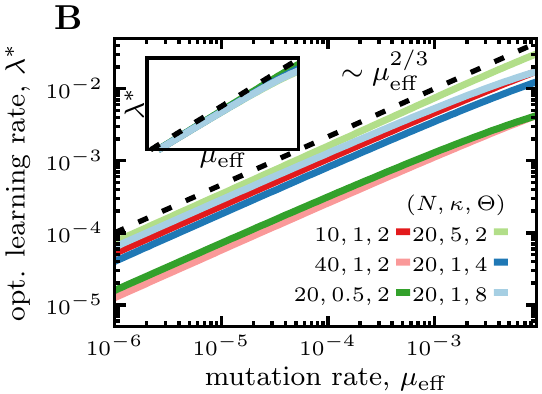}
\includegraphics[width= 0.33\textwidth]{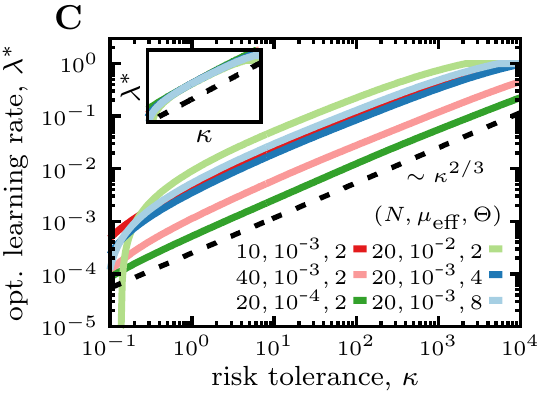} 
 \caption{ \textbf{Optimal memory strategy for evolving patterns.} {\bf (A)} The objective function  $\Q(\lambda)$ (Eq.~\ref{eq:Objective}) is shown as a function of the learning rate,  for $\kappa=1$ and $\Theta =2$. Analytical approximations (dashed lines) are compared to numerical estimates (full lines), for different mutation rates $\mu_\eff$ (colors).  {\bf (B, C)}  The optimal learning rate (Eq.~\ref{eq:opt_lambda}) is shown as a function of the mutation rate $\mu_\eff$ (B) and the risk tolerance $\kappa$ (C), for different sets of parameters indicated in the legend. Insets show collapsed plots after rescaling according to Eq.~\ref{eq:opt_lambda}. }
\label{fig:Fig1} 
\end{figure*}

To determine the memory performance for a given stimulus $\chi$, we define the affinity $\A(\M,\chi)$ as the overlap between the patterns stored in the memory repertoire $\M$ and the pattern $\chi$,
\begin{equation}
\label{eq:affinity}
\A (\M,\chi) = \A_0 \sum_{\psi^\alpha \in \M} m^\alpha |\braket{\psi^\alpha |\chi}|^\Theta  - { \A_{\rand}},
\end{equation}
Here, we used a short-hand notation to denote the normalized pattern vector by $\ket\psi \equiv \psi/\sqrt{L}$, its transpose by $\bra\psi$, and a  normalized scalar product  $\braket{\psi|\chi}\equiv \left(\sum_i\psi_i\chi_i\right)/L$. We use the shape parameter $\Theta$ to modulate the dependency of the affinity function on pattern overlap. The scaling parameter $\A_0$ is model dependent and sets the unit of affinity in a given system,  yet its precise value does not impact our analysis. The offset ${\A_{\rand}}$ is chosen such that the expected affinity of random patterns remains zero.  The choice of affinity as a measure of performance is inspired by memory retrieval in biological systems, where recognition is mediated through biophysical interactions.

We characterize the response of a memory repertoire to $N$ independently evolving pattern classes $\Psi^c$ (with $c\in{1,\dots,N}$). Each  class $\Psi^c$ denotes a set $\{\psi\}_c$ of patterns generated over time through evolution. Patterns within each class  evolve (mutate) by random spin-flips of rate $\mu$.  At each time step $t$, a pattern from a randomly chose class is presented to the memory repertoire, resulting in an effective observed mutation rate of $\mu_\eff= N\mu$ per expected encounter with the same pattern class. To simplify, we use $\Psi^c(t)$ to denote the evolved pattern from class $c$ that is generated and presented to the repertoire at time $t$. Since pattern classes are  orthogonal to each other (up to finite-size effects), the expected overlap between patterns presented at different times follows $\av{\braket{\Psi^c(t)|\Psi^{c'}(t+\tau)}} =\delta_{c,c'} \rho^\tau +\mathcal{O}(1/\sqrt{L})$, where $\rho= (1-2\mu)$ measures the similarity between evolved patterns.

Interestingly, for $\Theta=2$ this model is equivalent to the energy function of the classical Hopfield network with  Hebbian learning rule~\cite{Hopfield:1982fq,workingMem87,Amit:1985bo,Schnaack:Hop1} (Appendix~\ref{sec:MapToHop}). This correspondence enables us to simulate the memory repertoire efficiently and to test analytic predictions with numerical experiments; see Appendix~\ref{sec:Nummet} for numerical method.

\section{Results}
\subsection{Optimal learning for evolving patterns with risk-return tradeoff}
 We seek to find an optimal  strategy to set the learning rate such that the stored memory in  a repertoire  can be reliably retrieved for patterns evolving at a specified rate $\mu$. Learning and updates of  memory repertoires over time  impact both the expected affinity and the variance of the affinity across patterns. A high learning rate can sustain a high affinity in a repertoire for the most recently presented patterns. However, keeping up with the latest trend can be risky as the variance of the affinity across patterns can increase, with older patterns suffering most from this tradeoff. To  account for this effect, we optimize an objective function $\Q(\lambda;\mu)$ that balances the risk-utility tradeoff  by maximizing the mean (utility) $\av\A$ and minimizing the standard deviation (risk) $\sigma_\A$ of the affinity across patterns,
\begin{equation}
\label{eq:Objective}
\Q(\lambda;\mu) = \av{\A } - \frac{1}{\kappa} \, \sigma_{{}_\A}.
\end{equation}  
Here, $\kappa$ measures the risk tolerance of the repertoire. Such a risk-utility  analysis was initially introduced in economics~\cite{steuer_multiple_1986,sen_markets_1993}, but has  since been used to characterize tradeoffs in  biological  and evolutionary processes~\cite{shoval_evolutionary_2012,schuetz_multidimensional_2012,hart_inferring_2015,szekely_mass-longevity_2015,seoane_phase_2015,tendler_evolutionary_2015,kocillari_signature_2018}. 

We can analytically evaluate both the mean and the variance of the affinity by using explicit expansion of the encounter history or by evaluating the cumulant-generating function for the distribution of affinities, valid  in the  large-$N$  limit (i.e, many patterns); see Appendix~\ref{sec:appStatsAffinity} and Fig.~\ref{fig:STD_patt_SI}. The expression for the $n^{th}$ cumulant of the affinity follows,
\begin{align}
\label{eq:cumulants}
c_n=a_0^n\frac{\lambda^n \rho^{\Theta n}}{N \left(1 - (1-\lambda)^n \rho^{\Theta n}\right)} +\mathcal O\left(N^{-2}\right)
\end{align} 
where $a_0= \A_0-\A_\rand$, and the mean and the variance are given by $n=1$ and $n=2$, respectively.  For static patterns  ($\mu = 0 \leftrightarrow \rho=1$), the mean affinity becomes independent of the learning rate and reaches the na\"ive expectation $a_0/N$. For evolving patterns, a repertoire with a maximal learning rate $\lambda=1$ can achieve  the highest mean  affinity $a_0 \rho^\Theta/N$ by storing a memory of the most recent pattern with the maximal affinity $a_0$, while treating the other patterns as random. 

\begin{figure}[h!]
\centering
\includegraphics[]{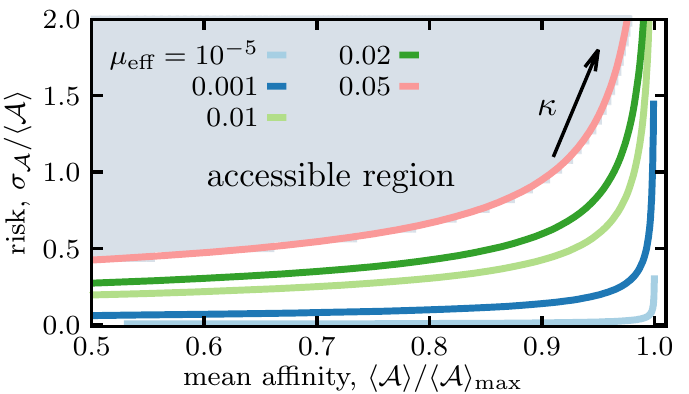}
 \caption{ {\bf Pareto front for risk-utility tradeoff of memory for evolving patterns.} Optimized objective function $\Q(\lambda^*)$ (Eq.~\ref{eq:Objective}) describes a Pareto front in  the scaled affinity-risk plane by varying the risk tolerance $\kappa$ along each line,  for different mutation rates (colors). To present the Pareto front in a dimensionless space,  the mean affinity is scaled by its maximal value $\langle\A\rangle_\text{max} =\frac{a_0}{N}$, and the standard deviation as risk is scaled by the mean affinity. Only the combinations of risk and affinity values that lie above the Pareto front (gray shading) are accessible to  a memory repertoire.
 Other parameters: $N=200$ and $\Theta =2$. Results for other shape parameters are shown in Fig.~\ref{fig:Pareto_SI}.}
\label{fig:Fig2} 
\end{figure}

For a broad range of evolutionary rates, the optimum of the objective function  $\Q(\lambda;\mu)$ is achieved for intermediate values of  learning rate $\lambda$, where the mean and variance of affinity across patterns are balanced; see Fig.~\ref{fig:Fig1}A for analytical and numerical results. The optimal learning rate  $\lambda^* (\mu)= \max_\lambda \Q(\lambda;\mu)$ that maximizes the objective function scales with  different model parameters as,
\begin{equation}
\label{eq:opt_lambda}
\lambda^* = \frac{2}{N}\left(2 \kappa \, \Theta\, \mu_\eff\right)^{2/3} + \mathcal O\left(\frac{\mu_\eff}{N} \right).
\end{equation}
The analytical scaling relation  in Eq.~\ref{eq:opt_lambda} is in excellent agreement with numerical simulations (Fig.~\ref{fig:Fig1}B,C). As the evolutionary rate of patterns $\mu_\eff$  increases, the optimal learning rate grows so that the repertoire closely follows the patterns' evolution (Fig.~\ref{fig:Fig1}B).  As the shape parameter $\Theta$ increases, the affinity function becomes more  peaked around the recently stored patterns (Eq.~\ref{eq:affinity}), resulting in an increase in the optimal learning rate to keep the memory  focused on the more recent (less evolved) encounters.  The learning rate scales inversely with the number of patterns $N$ for the repertoire to evenly distribute the resources (allocate memory).  Repertoires with higher risk tolerance $\kappa$  use larger learning rates (Fig.~\ref{fig:Fig1}C) and store a risky but a high-affinity memory against recent encounters. As  repertoire becomes more risk-avert (small $\kappa$), it  stores a more equitable memory across patterns but at a loss for affinity.  In the limit of no risk tolerance ($\kappa \rightarrow 0$), the repertoire stops learning ($\lambda = 0$) and adopts a risk-free but impractical strategy where the memory has zero affinity for all patterns. This tradeoff can be depicted by a Pareto front in the affinity-risk space, along which one cannot increase the mean affinity without increasing the risk or vice versa. Fig.~\ref{fig:Fig2} shows these Pareto fronts parametrized by scaled mean affinity and risk,  for  different mutation rates (colors) and by varying the risk tolerance $\kappa$ along each line. The combinations of risk and affinity values that lie below the Pareto front are inaccessible, and those that lie  above are sub-optimal solutions for a memory repertoire.

\subsection{Discrimination of random and stored patterns}
One key goal of memory usage is to recognize and accurately classify presented patterns with a memory stored from prior encounters. A misclassification  could have dire consequences. For example, in the case of immune system, if memory response is not triggered by a secondary infection (i.e., \emph{false negative}), the host would pay a cost by enduring sickness and having to mount a novel response and re-store a new memory. \emph{False positive} responses are also costly, as they can  be associated with autoimmunity if mounted against self-antigens~\cite{AltanBonnet:2020hk}, or they can interfere with novel responses without preventing the disease, e.g., in the case of original antigenic sin against viruses like influenza~\cite{cobey_immune_2017}.    

Memory strategies optimized to operate under different risk tolerance $\kappa$ can yield varying levels of pattern misclassification. We characterize the discrimination accuracy of a repertoire by quantifying the rate by which it recognizes evolved patterns associated with previously stored memory (true positive), or  randomly generated patterns without any prior encounter history (false positive). To do so, we need to characterize the distribution of affinities for patterns with prior encounter history, and  for novel (random) patterns.

The recognition affinity of a memory repertoire for recurring patterns   depends on the history of pattern encounters and on the learning rate $\lambda$. For example, in the case where  patterns from a given class $\Psi^c$ are presented at time points $[t_1, \dots, t_n]$, the affinity $\A(\Psi^c(T))$ for an evolved pattern  from the same class shown at time $T>t_n$ can be expressed as
\begin{align}
\A(\Psi^c (T)) = a_0\sum_{i=1}^n \lambda e^{-\lambda(T- t_i -1)} \rho^{\Theta (T- t_i)} \label{Eq.affinity}
\end{align}
Here, the factor $e^{-\lambda(T- t_i-1)}$ accounts for the exponential decay for the affinity of a memory  stored at time $t_i$  from its maximum level $\lambda$, due to updates in the repertoire (Eq.~\ref{eq:decay_time}).   $\rho^{\Theta (T- t_i)} $ accounts for the decay in the overlap between the memory stored at time $t_i$ and the presented patterns in the future, due to evolution. 

Patterns are presented to the repertoire in random order, and  the time  $\tau_i= t_i - t_{i-1}$ between consecutive encounters with the same class  is exponentially distributed with a mean of $N$ steps (i.e., number of classes), $p(\tau)\approx e^{-N\tau}/N$. The distribution of affinities $P_\Psi(\A)$ for patterns of a given class can be derived by convolving the affinity function in Eq.~\ref{Eq.affinity} with the exponential waiting time distribution for pattern history. Although the exact form of this affinity distribution  is difficult to evaluate, we expect it to be from an exponential family. Indeed, the Gamma distribution is a good approximation to the distribution of the affinities  (Fig.~\ref{figHistogram_SI}). We  quantify the accuracy  of this approximation by the Kullback-Leibler distance $D_{\mathrm{KL}}(P_\Psi(\A)\| \Gamma_\A)$ between the true affinity distribution $P_\Psi(\A)$ and a Gamma distribution $\Gamma_\A$ with the same mean and variance. By using an Edgeworth expansion~\cite{hall_bootstrap_1992,blinnikov_expansions_1998} with the cumulants of the affinity distribution  in Eq.~\ref{eq:cumulants}, we can  show that  in the limit of small learning and mutation rates the Kullback-Leibler distance  between the two distributions is small, with $D_{\mathrm{KL}}(P_\Psi(\A)\|\Gamma(\A)) = (25/27) \left(2 \Theta\mu_\eff/\kappa\right)^{2/3} + \mathcal{O}(\Theta \mu_\eff) \ll 1$ (Appendix~\ref{subsec:Edge}). This result can also be intuitively understood, since the Gamma distribution arises  in processes for which the waiting times between events are relevant. 

The distribution of affinities for random patterns (i.e., not belonging to any of the presented classes) $P_0(\A)$ can be similarly characterized.  The law of large numbers suggest that the overlap between unrelated patterns should be normally distributed with mean zero and variance $1/(4L)$. The overlap between the  memory repertoire and a random pattern is the sum of the normally distributed random overlaps weighed by the   weights $\{m^\alpha\}$,  associated with the stored patterns  (Appendix~\ref{subsec:Affinity_rand} and Fig.~\ref{fig:STD_Rand_SI}). Thus, the distribution of affinities for random patterns $P_0(\A)$  is well-approximated by a Gamma distribution with mean $\propto L^{-\Theta/2}$ and variance $\propto L^{-\Theta}$. This distribution should be contrasted to that of the affinities for patterns with prior encounter histories $P_\Psi(\A)$, the statistics of which primarily depends  on the  number of patterns and the learning rate, with mean $\propto  1/N$, and variance $\propto  \lambda/N$.
\begin{figure}[]
\centering
\includegraphics[]{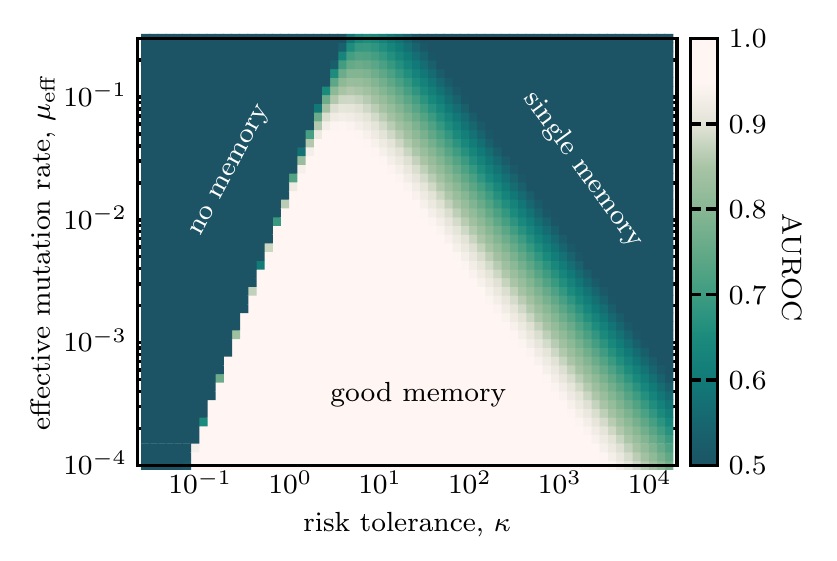}
 \caption{ \textbf{Three phases of memory discrimination.} 
The phase diagram shows the discrimination ability of repertoires (AUROC) between familiar patterns with prior encounter history  and random patterns, based on their respective affinities. AUROC is estimated using numerical approximations to the affinity distributions $P_\Psi(\A)$ and $P_0(\A)$ for familiar  and random patterns (Appendix~\ref{sec:appStatsAffinity}).  Each point in the phase diagram shows  AUROC for a repertoire optimized with a given risk tolerance $\kappa$ for  patterns with a specified  mutation rate $\mu_\eff$. Other parameters: $L=200$, $N=40$, and $\Theta =2$.  Results for other shape parameters $\Theta$ are shown in Fig.~\ref{fig:ROC_SI}. Full simulation results are shown in Fig.~\ref{fig:ROC_DATA_SI}.}
\label{fig:Fig3} 
\end{figure}

The sensitivity (fraction of true positives to all patterns associated with memory) and specificity (fraction of false negatives to all random patterns) of a repertoire in classifying patterns depends on an affinity cut-off for distinguishing between familiar and random patterns. A receiver operating characteristic (ROC) curve  shows the relationship between sensitivity and specificity of a classification in a repertoire for every possible affinity cut-off. The  area under the ROC curve (AUROC), which measures the discriminative ability of the repertoire, depends on the evolutionary rate  $\mu_\eff$ of patterns it encounters, its  risk tolerance $\kappa$, and the parameter   $\Theta $ that determines the shape of the affinity function (Eq.~\ref{eq:cumulants}). These parameters also determine the optimal learning rate $\lambda^*$ (Eq.~\ref{eq:opt_lambda}), which sets the memory strategy of a repertoire in a given evolutionary setup.

The phase diagram for the  discrimination ability of a repertoire defines three distinct regions determined by a combination of  risk tolerance  $\kappa$ and the evolutionary rate of patterns $\mu_\eff$, for a specified shape parameter $\Theta$ (Fig.~\ref{fig:Fig3}): (i) A triangular region  at the center of the phase diagram with $\text{AUROC} \simeq 1$ indicates  the range of parameters for which the repertoire can efficiently discriminate between familiar and novel patterns. Within this region, as the evolutionary rate of patterns increases, the range for risk tolerance that enables a repertoire to efficiently discriminate between familiar and random patterns narrows, and the repertoire has to fine-tune its learning rate to match the faster evolution of the patterns (Fig.~\ref{fig:Optimal_learning_heat_SI}). (ii) When risk tolerance $\kappa$ is  large, the repertoire updates rapidly and  only keeps a memory of  the most recent encounters, resulting in an inefficient discrimination between familiar and novel patterns with $\text{AUROC} \simeq 0.5$. (iii) For risk avert strategies (small $\kappa$), the memory  effectively shuts down and  the repertoire  cannot anymore discriminate between random and familiar patterns.  It should be noted that a change in  the affinity shape parameter $\Theta$  can shift the exact  boundaries between these phases, but it does not impact the overall structure of the phase diagram (Fig.~\ref{fig:ROC_SI}).

Taken together, a moderate  risk tolerance ($\kappa \Theta = \mathcal{O}(1)$) enables a repertoire to store an effective and a robust memory and to operate without fine tuning. We further confirm these results with simulations for $\Theta=2$ in Fig.~\ref{fig:ROC_DATA_SI}.
\section{Discussion}
Devising a learning strategy to store functional memory for evolving stimuli is an open problem with potential applications in many fields. Inspired by immune memory that  reliably recognizes  evolving threats, we present an analytical energy-based memory model against evolving patterns that captures the risk-utility tradeoff in memory repertoires. 

We found that risk-tolerant repertoires adopt faster learning rates and keep a memory of the recent patterns with high affinity,  at the risk of disregarding older patterns.  On the other hand, risk-avert repertoires effectively shut down their learning to minimize the variance in their recognition affinity for different patterns, at the cost of having a low affinity for all patterns. A moderate risk tolerance enables a repertoire to achieve a desirable balance between risk and affinity (utility) to robustly classify evolving patterns and associate them with prior memory. This risk-utility tradeoff defines Pareto front for accessible memory strategies  by a repertoire (Fig.~\ref{fig:Fig2}). 

One key role of memory storage is to reliably discriminate between familiar and novel stimuli. The discrimination ability of an optimal memory repertoire depends on its risk tolerance and  the  evolutionary rate of the patterns that it encounters.  Interestingly, as the evolutionary rate of the presented patterns increases, the range of risk tolerance that allows a repertoire to retain a functional memory narrows and approaches a regime where risk and utility are equally valued. The classification criteria in our analysis are solely based on the affinity of the memory repertoire to a presented pattern.  This stands in contrast to energy-based Hopfield-like networks that achieve recognition by retrieving associative memory stored in the networks' energy minima~\cite{Hopfield:1982fq}.

We have used the fluctuations (standard deviation) of the memory's affinity over the ensemble of presented patterns to measure the risk of misclassification by the stored storage. While variance might be more suitable in other settings \cite{schweizer_meanvariance_1992,toft_meanvariance_1996}, using standard deviation as a measure of risk keeps the risk tolerance $\kappa$ dimensionless and comparable across different systems. Nonetheless, we expect the Pareto front's overall structure for the risk-utility tradeoff and the phase diagram for discrimination ability of the repertoires to remain qualitatively intact, irrespective of the exact choices made for the risk function.

Although our model is inspired by immune memory repertoires, the introduced approach is general enough that can be applied to other models of memory. In particular, incorporating risk and utility in training of artificial neural networks can guide these machine learning efforts to store efficient of memory for evolving signals~\cite{ditzler_learning_2015}.

\section*{Acknowledgement}
This work has been supported by the NSF CAREER award (grant No: 2045054), DFG grant (SFB1310) for Predictability in Evolution and the MPRG funding through the Max Planck Society. O.H.S also acknowledges funding from Georg-August University School of Science (GAUSS) and the Fulbright foundation.

\newpage

\onecolumngrid
\appendix
\newpage
  \setcounter{figure}{0}
 \renewcommand{\thefigure}{S\arabic{figure}}
\renewcommand{\theequation}{S\arabic{equation}} 

\section{Statistics of the affinity distribution}
\label{sec:appStatsAffinity}
 \subsection{The average affinity of a repertoire for presented patterns}
 In the main text (Eq.~\ref{Eq.affinity}) we express the affinity of a pattern class at time $T$ in terms of the  prior encounters with the patterns of the same class at times $[t_1,...,t_n]$ with $t_{n-1}<t_n<T$.  To compute the statistics  (mean and variance) of a repertoire's  affinity, we first express the affinity function in terms of the  times $\hat{t}_i = T - t_{n-i+1}$, passed since the  $i^{th}$ encounter, when counting backward in time. This allows us to write the affinity as 
 \begin{align}
 \label{eq:affinity_revers}
 \A(\Psi^c(T))= a_0 \lambda\sum_{i=1}^n  e^{-\lambda \hat{t}_i-1} \rho^{2  \hat{t}_i}.
 \end{align}
Similar to forward times $t_i$, the reverse  times $\hat{t}_i$ are separated by exponentially distributed independent waiting times $\hat{\tau}_i = \hat{t}_{i} - \hat{t}_{i-1}$,   
\begin{align}
\label{eq:tau_hat_dist}
p(\hat \tau) = \frac{(N-1)^{ \hat\tau-1}}{N^{\hat{\tau}} } \approx \frac{1}{N} e^{-N  \tau}.
\end{align}
Given the relationship $\sum_{j=1}^i \hat{\tau}_j = \hat{t}_i$, we can express the affinity of the patterns in Eq.~\ref{eq:affinity_revers} in terms of the statistically independent $\hat{\tau}_i$ as,
  \begin{align}
 \label{eq:Affinty_full_rand}
 \A(\Psi) &=a_0 \lambda  \sum_{g=1}^n (1-\lambda)^{\sum_{g'=1}^g \hat{\tau}_{g'} -1} \rho^{\Theta \sum_{g'=1}^g \hat{\tau}_{g'}}.
 \end{align}
Here, we can interpret each term in the sum as  the contribution of a memory from the corresponding prior generation to the affinity at the current time. These contributions decay due to the updates (learning) in the repertoire with rate $(1-\lambda)$ and due to evolution of the patterns with rate $\rho$. The expected  affinity contribution $\langle \A_g\rangle$  from the $g^{th}$ generation follows, 
\begin{align}
\nonumber\langle \A_g\rangle &= a_0 \lambda \sum_{\hat\tau_1} \dots \sum_{\hat\tau_{g}}\left( \prod_{i=1}^g p(\hat\tau_{i}) \right)(1-\lambda)^{\sum_{i=1}^g \hat{\tau}_{i} -1} \rho^{\Theta \sum_{i=1}^g \hat{\tau}_{i}}\\
\nonumber & = a_0 \lambda (1-\lambda)^{-1} \prod_{i=1}^g \left(\sum_{\hat\tau_{i}} p(\hat\tau_{i}) (1-\lambda)^{\hat\tau_{i}}   \rho^{\Theta \hat\tau_{i}}\right)\\
\label{eq:mean_affinity_gen_g}
& = a_0 \lambda (1-\lambda)^{-1} \left(\frac{(1-\lambda) \rho^\Theta}{N +(1-\lambda) \rho^\Theta -  N(1-\lambda) \rho^\Theta  } \right)^g
\end{align}
where we used the fact that the time  windows $\hat{\tau}_i$ are independent from each other. The expected affinity follows from adding up the contributions from all generations,
 \begin{align}
\label{eq:SI_expected_affinity}
\langle \A\rangle =  \sum_{g=1}^\infty \langle \A_g\rangle =a_0\frac{ \lambda \rho^\Theta}{N  \left( 1- (1-\lambda)\rho^\Theta \right) }.
\end{align}

As mentioned in the main text, this result immediately shows that the system reaches the maximal mean affinity of $a_0/N \rho^\Theta$ for the maximal learning rate $\lambda =1$. Moreover, the mean affinity becomes independent of the learning rate when the patterns are static ($\rho=1$).

\subsection{The variance of repertoire's affinity across presented patterns}

To calculate the  variance of the affinity (Eq.~\ref{eq:mean_affinity_gen_g} in the main text) we need to account for the covariance {$\cov (\A_{g_1} , \A_{g_2})$} between the  contributions from different generations. Because the covariance is symmetric, we can write the variance of the affinity as 
\begin{align}
\label{eq:Def_varA}
\var\left(\A \right) = \sum_{g=1}^\infty \var( \A_g ) + 2 \sum_{g_1=1}^\infty\sum_{g_2=g_1+ 1}^\infty \cov \left( \A_{g_1},\A_{g_2}\right).
\end{align} 

First we calculate the variance of the individual terms $\A_g$. Following the notation introduced in Eq.~\ref{eq:mean_affinity_gen_g}, we find
\begin{align}
\nonumber
\var(\A_g)  =& \var \left( a_0 \lambda (1-\lambda)^{-1} \prod_{i=1}^g \left((1-\lambda)^{\hat\tau_{i}}   \rho^{\Theta \hat\tau_{i}}\right) \right) \\
=& 
\label{eq:var_A_g_line2}
 \var \left( a_0 \lambda (1-\lambda)^{-1} \prod_{i=1}^{g-1} \left((1-\lambda)^{\hat\tau_{i}}   \rho^{\Theta \hat\tau_{i}}\right)\ \left((1-\lambda)^{\hat\tau_{g}}   \rho^{\Theta \hat\tau_{g}}\right) \right)  \\
\nonumber 
=& \left\langle (1-\lambda)^{\hat\tau_{g}}   \rho^{\Theta \hat\tau_{g}}  \right\rangle^2  \var \left( a_0 \lambda (1-\lambda)^{-1} \prod_{i=1}^{g-1} \left( (1-\lambda)^{\hat\tau_{i}}   \rho^{\Theta \hat\tau_{i}}\right) \right) \\
\label{eq:var_A_g_line3}
&+ \left\langle a_0 \lambda (1-\lambda)^{-1} \prod_{i=1}^{g-1} \left((1-\lambda)^{\hat\tau_{i}}   \rho^{\Theta \hat\tau_{i}}\right)  \right\rangle^2  \var  \left( (1-\lambda)^{\hat\tau_{g}}   \rho^{\Theta \hat\tau_{g}}  \right) \\ 
\label{eq:var_A_g_line4}
=&\left( a_0 \lambda (1-\lambda)^{-1}  \right)^{-2} \left(\left\langle \A_1\right\rangle^2 \var\left( \A_{g-1}\right) + \left\langle \A_{g-1} \right\rangle^2 \var\left(\A_{1}\right) \right) \\
\label{eq:var_Ag_final}
=& g\left( a_0 \lambda (1-\lambda)^{-1}  \right)^{-(2g-2)}   \left\langle\A_1 \right\rangle^{2g-2} \var\left(\A_1\right). 
\end{align}
Because the time intervals $\hat{\tau}_i$ are independent from each other, we could use error propagation to get from Eq.~\ref{eq:var_A_g_line2} to Eq.~\ref{eq:var_A_g_line3}. Moreover, since all time time intervals follow the same  distribution (Eq.~\ref{eq:tau_hat_dist}), we could insert the statistics of $\A_g$ with the correct normalization $\left( a_0 \lambda (1-\lambda)^{-1}  \right)^{-2} $ to arrive at Eq.~\ref{eq:var_A_g_line4}. Finally, we performed $g-1$ further iterations to get a result that only depends on the statistics of the first contribution $\A_1$ in~Eq.~\ref{eq:var_Ag_final}. Thus, it is  sufficient to calculate the variance of only one generation,
\begin{align}
\label{eq:Variance_D1}
\var(\A_1) =a_0^2 \rho ^{2\Theta} \left(\frac{1}{N-(1-\lambda )^2 (N-1) \rho ^{2\Theta}}-\frac{1}{\left(N-(1-\lambda ) (N-1)
   \rho^\Theta\right)^2}\right). 
\end{align}
To evaluate the variance of the affinity (Eq.~\ref{eq:Def_varA}), we still need to calculate the covariance between the  contributions from different generations. Similar to Eq.~\ref{eq:var_A_g_line2}, we use, 
\begin{align}
\A_g = \A_{g-i} \cdot \prod_{j=g-i}^g \left((1-\lambda)^{\hat\tau_{j}}   \rho^{\Theta \hat\tau_{j}}\right).
\end{align} 
which entails the following relationship for the covariance between $\A_{g}$ and $\A_{g-i}$ , 
{\small \begin{align}
\nonumber
\cov\left(\A_{g},\A_{g-i}\right) &=\cov\left(\A_{g-i} \cdot \prod_{j=g-i}^g \left((1-\lambda)^{\hat\tau_{j}}   \rho^{\Theta \hat\tau_{j}}\right),\A_{g-i}\right)  = \var(\A_{g-i}) \left\langle\prod_{j=g-i}^g \left((1-\lambda)^{\hat\tau_{j}}   \rho^{\Theta \hat\tau_{j}}\right).\right\rangle \\ 
\label{eq:cov_A1_A2}
&= \left( a_0 \lambda (1-\lambda)^{-i}  \right)^{-1}\var(\A_{g-i}) \langle\A_1\rangle^{i}.
\end{align}}
Here, we  again used the fact that all time intervals are independent and that they  follow the same distribution in Eq.~\ref{eq:tau_hat_dist}.
Using the expressions for variance of individual terms (Eq.~\ref{eq:var_Ag_final}) and the covariance between them (Eq.~\ref{eq:cov_A1_A2}), we can characterize the variance of the affinity (Eq.~\ref{eq:Def_varA}) as,
\begin{align}
\label{eq:theory_variance_energy}
\var(\A) = a_0^2 \left[ \frac{\lambda  (N-1) \rho ^{2\Theta} \left(N-(1-\lambda ) (N-1) \rho^{\Theta}\right)^2 \left(\lambda  N-(1-\lambda ) \rho^{\Theta} (\lambda  (N+2)-2)\right)}{N^2 \left(1-(1-\lambda ) \rho^{\Theta}\right) \left(N-(1-\lambda ) (N-2)  \rho^{\Theta}\right)^2 \left(N-(1-\lambda )^2 (N-1) \rho ^{2\Theta}\right)}  \right].
\end{align}
Note that the expression in Eq.~\ref{eq:theory_variance_energy} is accurate up to terms of order $a_0^2 \mathcal O(N^{-1}L^{-\Theta})$, which arise from the (negligible) overlap between patterns of difference classes due to the finite size of the patterns; see Appendix~\ref{subsec:Affinity_rand}.

\subsection{Cumulant-generating function for the affinity distribution}
To characterize the distribution of the affinities, we can rely on the corresponding cumulant-generating function. The cumulant-generating function $R_X(q)$ of a random variable $X$ is defined as the logarithm of the moment-generating function 
\begin{align}
\label{eq:def_cumlantG}
R_{X}(q) = \log \av{e^{q X}}.
\end{align}
If we define a new random variable $Y  = \sum_i X_i$ as the sum of independent random variables, its  cumulant-generating function is given by $\R_Y(q) = \sum_i R_{X_i}(q) $.  To use this relation for the affinity function, we need to rewrite the affinity of a pattern class (Eq.~\ref{eq:affinity_revers}) as a sum of independent random variables. In Eq.~\ref{eq:affinity_revers} we calculate the affinity as the sum over the contributions of past encounters. In that view, the times of the past encounters $\hat{t}_i$ are random variables while the contributions to the affinity at  those times are deterministic. 

We now change the point of  view and write the affinity function as sum of contributions from all time, instead of only the past encounters (Eq.~\ref{eq:affinity_revers}),
\begin{align}
\label{eq:AsumOverTime}
\A (\Psi^c) = \sum_{\hat{t}=1} \A^c( \Psi^{c'}(\hat{t}))
\end{align} 
Here, $\hat{t} $ is the reverse time with the origin at the final (current) time point, as defined in  Eq.~\ref{eq:affinity_revers}. In this case, the  encounter of a network with pattern  $\Psi^{c'}(\hat{t})$  from class $c'$  at (reverse) time $\hat{t}$   is accounted for by the contribution $\A^c( \Psi^{c'}(\hat{t}))$ to the network's affinity against the pattern $\Psi^c$. We then find 
\EQ
\A^c( \Psi^{c'}(\hat{t})) =
 \begin{cases}
 \label{eq:SI_affinityContrib}
 0 & \text{if } c\neq c'\\\\
a_0 \lambda (1-\lambda)^{\hat{t}-1 }  \rho^{\Theta\, {\hat{t}}} =:  E(\hat t) & \text{if } c=c'
 \end{cases}
 \EE
which implies that a repertoire's affinity against a pattern from a given class is only determined by the prior history of the repertoire's encounters with patterns of the same class, and the memory of these  prior encounters decay over time  according to Eq.~\ref{eq:affinity_revers}. 

The repertoire encounters a pattern of a specific  class at a given time point  with probably $1/N$. As a result, the distribution of affinity contributions from a given time point $\A^c( \Psi^{c'}(\hat{t}))$ can be expressed as,
\begin{align}
\label{eq:SI_affinty_dist_at_t}
P[\A^c( \Psi^{c'}(\hat{t}))]= \left(1-\frac{1}{N}\right) \, \delta\left(\A^c( \Psi^{c'}(\hat{t})) -0\right) + \frac{1}{N}\, \delta\left(\A^c( \Psi^{c'}(\hat{t}))- E(\hat{t})\right),
\end{align}
where $E(\hat t)$ is defined in Eq.~\ref{eq:SI_affinityContrib}. As a result, the expectation value $\av{\A (\Psi^c(T)) }$ for the  affinity of the repertoire against a pattern from class $c$  (Eq.~\ref{eq:AsumOverTime}) can be evaluated as,  
\begin{align}
\label{eq:mean_A_second_version}
\av{\A(\Psi^c)} = \sum_{\hat{t}=1} \int
\dd \A^{c,c'}_{\hat{t}}   \A^{c,c'}_{\hat{t}} P[ \A^{c,c'}_{\hat{t}}]. 
\end{align} 
where we used the shorthand notation  $ \A^{c,c'}_{\hat{t}} \equiv  \A^c( \Psi^{c'}(\hat{t}))$.
Fig.~\ref{figHistogram_SI} shows an agreement between simulations and the expected affinities estimated with this procedure.

We can now express the cumulant-generating function of the affinity distribution  $P[\A^c( \Psi^{c'}(\hat{t}))]$ as the sum of the cumulant-generating functions of the independent terms $\A^c( \Psi^{c'}(\hat{t})$. With the definition in Eq.~\ref{eq:def_cumlantG} we can evaluate the cumulant-generating functions $R_{\hat{t}}(q,\hat{t})$ for the contributions of each time point $\A^c( \Psi^{c'}(\hat{t})$ to the affinity function,
\begin{equation}
\begin{split}
R_{\hat{t}}(q,\hat{t})= \log \av{e^{q \A_{\hat{t}}(\hat{t}) }} &= \log \left( (1 - \frac{1}{N})   + \frac{1}{N}  e^{ q E(\hat{t})} \right) \\
\label{eq:cumulant_G_atT}
 &\approx - \frac{1}{N} + \frac{1}{N} e^{ q E(\hat{t})} + \mathcal O\left(N^{-2}\right).
 \end{split}
\end{equation}

The cumulant-generating function $\R(q)$ of the affinity distribution  $P[\A]$ can be expressed as  the sum of the cumulant-generating functions of the independent contributions from each time point (Eq.~\ref{eq:cumulant_G_atT}), which entail, 
\begin{equation}
\begin{split}
\R(q) =\sum_{\hat t=1} R(q,\hat{t}) &=  \sum_{\hat t=1} \left[- \frac{1}{N} + \frac{1}{N} e^{ q E(\hat{t})}\right] + \mathcal O\left(N^{-2}\right)\\
\label{eq:CumulantsG_1}
&= \sum_{\hat t=1} \left[ - \frac{1}{N} + \frac{1}{N} \sum_{n=0} \frac{q^n E(\hat{t})^n}{n!}\right] +\mathcal O\left(N^{-2}\right)\\
&= \frac{1}{N}\sum_{n=1}\frac{q^n}{n!}a_0^n \sum_{\hat{t}=0} \lambda^n\rho^{\Theta n} \left((1-\lambda)\rho^\Theta \right)^{n \hat{t}} + \mathcal O\left(N^{-2}\right)\\
& =\frac{1}{N} \sum_{n=1}\frac{q^n}{n!} \times \frac{a_0^n  \lambda^n \rho^{n\Theta}}{1 - (1-\lambda)^n\rho^{\Theta n}} +\mathcal O\left(N^{-2}\right).
 \end{split}
\end{equation}
where we have substituted the exponential function with an infinite sum, used the expression in Eq.~\ref{eq:SI_affinityContrib} for  $E(\hat{t})$, and performed the resulting geometric sum.

We can now evaluate the $n^\text{th}$ cumulant $c_n$ of the affinity function as, 
 \begin{align}
 \label{eq:cumulantsSI}
 c_n = \left .\frac{\dd^n \R(q)}{\dd q^n}\right |_{q=0}  = a_0^n  \frac{\lambda^n \rho^{n\Theta}}{N\left(1 - (1-\lambda)^n\rho^{\Theta n}\right)}+ \mathcal O\left(N^{-2}\right).
 \end{align}

While the first cumulant is equal to the mean affinity in Eq.~\ref{eq:SI_expected_affinity}, obtained from the direct calculation of the moments, the second cumulant differs from the variance  in Eq.~\ref{eq:theory_variance_energy}. These differences arise due to the expansion of the logarithm in Eq.~\ref{eq:cumulant_G_atT}, which is only valid for large $N$. However, both results describe well the behavior of the variance   in the  regime that we are interested in (Fig.~\ref{fig:STD_patt_SI}), and therefore, it is warranted to use the simplified form in Eq.~\ref{eq:affinity} for our analysis. 

To characterize the scaling relation between the cumulants $c_n$ and the  model parameters, we can  expand Eq.~\ref{eq:cumulantsSI} for the case where the learning rate is close to its optimal value $\lambda^* \sim \mu_\eff^{2/3}$ (Eq.~\ref{eq:opt_lambda}), which entails, 
\begin{align}
\label{eq:cumulantsExpantion}
 c_n = a_0^n \frac{\lambda^{n-1}}{n N} + \mathcal O\left( \frac{\lambda^{n - 2} \mu_\eff }{n N^2} \right),
\end{align}
and thus, to the leading order, the mean affinity  scales as $c_1 \sim a_0 / N$ and the variance scales as $c_2\sim a_0^2 \lambda/(2 N)$.

\subsection{Approximating the affinity distribution with a Gamma distribution}
\label{subsec:Edge}
As mentioned in the main text, the interpretation of the affinity as a sum of samples from an exponential function separated by exponentially distributed times motivates us to model the affinity distribution as a $\Gamma$-distribution.  To corroborate this choice, we evaluate the  Kullback-Leibler divergence ($D_\KL$) between the distribution of patterns affinities in the repertoire $P_\Psi(\A)$ and a Gamma distribution with matching mean and variance. Since we do not have an analytical expression for $P_\Psi(\A)$  we use an Edgeworth approximation~\cite{blinnikov_expansions_1998} to evaluate the Kullback-Leibler divergence between the two distribution, by relying on the cumulants of $P_\Psi(\A)$ (Eq.~\ref{eq:cumulantsSI}). 

In brief, Edgeworth series expands a probability density function around a normal distribution in terms of its cumulants, and it provides a true asymptotic expansion with controlled error~\cite{blinnikov_expansions_1998}. To use the Edgeworth series, we transform the data to have mean zero and  variance one, resulting in a modified probability density function for affinities $\hat P_\Psi(\A)$. The second leading order approximation to  $\hat P_\Psi(\A)$ is  given by 
\begin{equation}
\label{eq:Edgeworth}
\hat P_\Psi(\A) \approx \phi(\A) \left( 1 + \frac{1}{3!} \hat c_3 \mathit{He}_3(\A) + \frac{1}{4!} \hat c_4 \mathit{He}_4(\A) + \frac{10}{6!} \hat c_3^2 \mathit{He}_6(\A) \right) := \phi(\A) \left( 1 + u_{P_\Psi} \right),
\end{equation}
where $\phi(\A) $ is a standard normal distribution, $\mathit{He}_n$ is a Hermite polynomial of order $n$,  and  $\hat c_n = c_n/c_2^{n/2}$  is  the $n^{th}$ normalized cumulant~\cite{blinnikov_expansions_1998,hall_bootstrap_1992}. As seen in Fig.~\ref{figHistogram_SI}, the approximation in Eq.~\ref{eq:Edgeworth} describes the distribution and especially the bulk of the affinity distribution very well. 

To  evaluate the Kullback-Leibler divergence $D_{\KL}(\hat{P}_\Psi||\hat{\Gamma}_\Psi)$ between the modified affinity distribution $\hat P_\Psi(\A)$ and a Gamma distribution $\hat\Gamma$ with matching mean and variance, we use an  Edgeworth expansion for both of these distributions,
\begin{align}
\nonumber D_{\KL}(\hat{P}_\Psi(\A)||\hat{\Gamma}_\Psi(\A)) &= \int_\A  \hat P(\A) \log \frac{  \hat P_\Psi(\A) }{ \hat \Gamma_\Psi(\A) } = \int_\A  \phi(\A) \left( 1 + u_{P_\Psi}\right)\log \frac{ \phi(\A) \left(1 +u_{P_\Psi}\right)}{\phi(\A) \left( 1 +u_{\Gamma_\Psi} \right)}\\
\label{eq:KL_edgworth_form2} &= \int_\A  \phi(\A) \left[ \left(  u_{P_\Psi} +\frac{1}{2} u_{P_\Psi} ^2\right) - \left(u_{\Gamma_\Psi}  + u_{P_\Psi} u_{\Gamma_\Psi}  - \frac{1}{2} u_{\Gamma_\Psi} ^2\right) \right] + \mathcal O(u^3)
\end{align}
where $u_{\Gamma_\Psi} $ arises from the Edgeworth expansion of the Gamma distribution, in analogy to Eq.~\ref{eq:Edgeworth}. A similar approach has previously been used to approximate the Kullback-Leibler divergence between two distributions~\cite{lin_edgeworth_1999,inglada_new_2007}.

To evaluate the integral in Eq.~\ref{eq:KL_edgworth_form2}, we use the orthogonality of the Hermit polynomials, i.e., 
\EQ\int_x \phi(x) \mathit{He}_n(x)\mathit{He}_m(x) = n! \delta_{n,m}\EE
 with $\mathit{He}_0=1$. As a result, all the linear terms in Eq.~\ref{eq:KL_edgworth_form2} vanish and  only the squared terms with equal polynomial orders $\mathit{He}_n$ contribute. We thus obtain,
\begin{align} 
\nonumber D_{\KL}(\hat{P}_{\Psi}(\A)||\hat{\Gamma}_{\Psi}(\A)) =& \frac{1}{2}\left(\frac{1}{3!} \hat c^2_3 +\frac{1}{4!} \hat c^2_4 +\frac{100}{6!} \hat c^4_3  \right) +\frac{1}{2}\left(\frac{1}{3!} \hat \gamma^2_3 +\frac{1}{4!} \hat \gamma^2_4 +\frac{100}{6!} \hat \gamma^4_3  \right) \\
\nonumber 
&- \left(\frac{1}{3!} \hat c_3 \hat\gamma_3 +\frac{1}{4!} \hat c_4 \hat\gamma_4+\frac{100}{6!} \hat c^2_3 \hat \gamma^2_3  \right) \\
\label{eq:KL_edgworth_final}
=&\frac{1}{2}\left[\frac{1}{3!} \left(\hat c_3 - \hat \gamma_3 \right)^2 +\frac{1}{4!} \left(\hat c_4 - \hat \gamma_4 \right)^2 +\frac{100}{6!} \left(\hat c_3^2 - \hat \gamma_3^2 \right)^2 \right] .
\end{align}
where $\hat \gamma_n$ is the $n^{th}$ cumulant of the modified Gamma distribution $\hat \Gamma_\Psi$. 

By substituting the cumulant of the affinity distribution (Eq.~\ref{eq:cumulantsSI}) and that of the Gamma distribution, we arrive at an approximation for the Kullback-Leibler distance between the two distributions, 
\begin{align}
\label{eq:SI_DKL_values}
D_{\KL}(P_{\Psi}(\A)||\Gamma_{\Psi}(\A)) = \frac{25}{27}  \left(2 \Theta \mu_\eff\kappa \right)^{2/3} + \mathcal O(\Theta \mu_\eff )\ll 1.
\end{align}
indicating that the  Gamma distribution is a good approximation for the true distribution of the  affinities $P_{\Psi}(\A)$ for small mutation rate $\mu_\eff$ as long as $\Theta \kappa$ is of order 1. However, for risk tolerant strategies (large $\kappa$), this approximation fails. In this regime the repertoire  only learns one pattern effectively, resulting in a bi-modal distribution of the affinities with one mode reflecting the low-affinity recognition and the other, the higher affinity of the latest stored pattern. This bimodal distribution cannot be approximated by a Gamma distribution.

\subsection{Affinity of random patterns}
\label{subsec:Affinity_rand}
The affinity of a random pattern $\chi$ (i.e., patterns unrelated to the previously encountered classes $\Psi^c$) is determined by  summing over the  overlaps $|\braket{\psi^\alpha|\chi}|^\Theta$ between the random pattern $\chi$ and all previously stored patterns in the memory repertoire $\psi^\alpha \in \M$, weighed by their contributions to memory $m^\alpha$. It should be noted that the mean affinity of random patterns is set to be zero and therefore, the affinity shift $\A_\rand$, defined in Eq.~\ref{eq:affinity},  can be evaluated as,
\begin{align}
\label{eq:DefArandSI1}
\A_\rand &= \av{\A_0\sum_{\psi^\alpha\in\M} m^\alpha |\braket{\psi^\alpha|\chi}|^\Theta} \\
\label{eq:DefArandSI2}
&= \A_0 \av{|\braket{\psi|\chi}|^\Theta} =\A_0 \frac{2^{\Theta/2} \Gamma\left[\frac{1+\Theta}{2} \right]}{\sqrt{\pi L^{\Theta}}},
\end{align}
where we use the fact that random pattern $\chi$ is independent of the stored patterns $\psi^\alpha$, and that on average their  overlap  $\braket{\psi|\chi}$ is a normally distributed random variable with mean zero and variance $\sigma^2_\rand = \frac{1}{4 L}$, according to the law of large numbers.

To evaluate the variance of random patterns, we first introduce a basis that spans all the stored patterns in a repertoire $\psi^\alpha$ that have non-zero weights $m^\alpha$ at a given time point $t$. One choice for this basis is to use the $N$ directions corresponding to the pattern classes $\Psi^c(t)$ at time $t$. However, these $N$ vectors do not fully span all the previously presented patterns, due to the evolutionary divergence of these patterns over time. To account for this remaining subspace, we introduce $N'$ auxiliary directions  $\Phi^{c}(t)$ (with $c'=1,\dots,N'$). In principle, the  space encompassing all possible the patterns is $L$-dimensional. However, the space of all stored patterns is typically more restricted (i.e., $N+N'<L$) due to the relatively  fast updates in the repertoire such that it only keeps a memory of  patterns that are similar to the current states of the presented classes $\Psi^c(t)$. Using the set of basis vectors $\{\Psi^c(t),\Phi^{c'}(t)\}$, we can express any stored pattern in the repertoire as, 
\begin{align}
\label{eq:DefBasis}
\bra{\psi^\alpha} = \sum_{c=1}^N \braket{\psi^\alpha|\Psi^c(t)  }\, \bra{\Psi^c(t)}+ \sum_{c'=1}^{N'} \braket{ \psi^\alpha|\Phi^{c'}(t) }\, \bra{\Phi^{c'}(t)}.
\end{align}
As a result, the overlaps between a random pattern $\chi$ and all the stored patterns in the repertoire in Eq.~\ref{eq:DefArandSI1} can be expressed as, 
{\small
\begin{align}
\label{eq:Bastransform1}
\sum_{\psi^\alpha\in\M} m^\alpha |\braket{\psi^\alpha|\chi}|^\Theta &= \sum_{\psi^\alpha\in\M} m^\alpha \left| \sum_{c=1}^N \braket{\Psi^c(t)|\chi} \, \braket{\psi^\alpha|\Psi^c(t)}+ \sum_{c'=1}^{N'} \braket{\Phi^{c'}(t)|\chi}\, \braket{\psi^\alpha|\Phi^{c'}(t)} \right|^\Theta \\
\label{eq:Bastransform2}
&\approx  \sum_{c=1}^N \left( \sum_{\psi^\alpha\in\M} m^\alpha \left|\braket{\psi^\alpha|\Psi^c(t)} \, \braket{\Psi^c(t)|\chi} \right|^{\Theta}\right) +  \sum_{c'=1}^{N'} \left( \sum_{\psi^\alpha\in\M} m^\alpha \left|\braket{\psi^\alpha|\Phi^{c'}(t)} \, \braket{\Phi^{c'}(t)|\chi} \right|^\Theta \right)\\
\label{eq:Bastransform3}
&\equiv \sum_{c=1}^N M_c\,  \left| {\braket{\Psi^c(t)|\chi}}\right|^{\Theta}  +
   \sum_{c'=1}^{N'} M'_{c'}\,  \left| {\braket{\Phi^{c'}(t)|\chi}}\right|^{\Theta}.
\end{align}}
To arrive at Eq.~\ref{eq:Bastransform2}, we assumed that the stored patterns $\psi^\alpha$ have non-vanishing overlaps with only one of the bases in the set $\{\Psi^c, \Phi^{c'}\}$.  As a result, in the expansion of the expression to the power $\Theta$ in Eq.~\ref{eq:Bastransform1}, all the cross terms  associated with different  bases  vanish, and the expression can be simply written as the sum of independent terms.  The final form in Eq.~\ref{eq:Bastransform3} expresses the overlap of a random pattern with the repertoire as a sum over the overlaps with the bases $\{\Psi^c(t),\Phi^{c'}(t)\}$, with effective weights $M_c(t)= \sum_{\alpha}m^\alpha \left|\braket{\psi^\alpha|\Psi^c(t)}\right |^\Theta $, and $M'_{c'}(t)= \sum_{\alpha}m^\alpha \left|\braket{\psi^\alpha|\Phi^{c'}(t)}\right |^\Theta $. The effective weight $M_c$ associated with the presented pattern classes $\Psi^c(t)$ follows,
\EQ
\label{eq:defMc}
M_c (t)= \sum_\alpha m^\alpha \left|\braket{\psi^\alpha|\Psi^c(t)} \right|^\Theta = \frac{\A\left(\Psi^c(t)\right) +\A_\rand }{\A_0}\approx \frac{\A\left(\Psi^c(t)\right) }{a_0},
\EE
where we used the approximation $a_0 = \A_0 - \A_\rand = \A_0+\mathcal{O}(L^{-\Theta/2})$.

When the shape parameter is $\Theta=2$, the effective weights define a normalized set (i.e., $\sum_{c=1}^N M_c + \sum_{c'=1}^{N'} M'_{c'}= 1$). However, for sharper affinity functions ($\Theta >2$), the relationship between the effective weights is bounded from above by   $\sum_{c=1}^N M_c + \sum_{c'=1}^{N'} M'_{c'}\leq 1$, and for broader affinity functions ($\Theta < 2$), the weights are bounded from below as, $1\leq\sum_{c=1}^N M_c + \sum_{c'=1}^{N'} M'_{c'}$. Here, we discuss the case with $\Theta >2$, which entails 
\EQ
\label{eq:DefSumMdash}\sum_{c'=1}^{N'} M'_{c'}(t) \leq 1-  \sum_{c=1}^N M_{c} (t)\approx 1- N\frac{\av{\A} }{a_0}.\EE
This upper bound implies that when the patterns are well represented in the repertoire (i.e., $\langle \A \rangle \approx  \A_{\max}  = a_0/N$; see Eq.~\ref{eq:SI_expected_affinity}), the weights of the auxiliary bases shrink to zero, i.e., $N' = 0$. This result is in line with the eigen-decomposition analysis of the generalized Hopfield network in Appendix~C of ref.~\cite{Schnaack:Hop1}. 

For the cases that $M'_{c'} >0$, we will consider a mean field approximation, where  all the auxiliary directions are equally important (i.e., $M'_{c'} = M', \, \forall c'$)  and that on average the memory stored in the auxiliary directions has a comparable weight to that of the bases spanned by the presented patterns, (i.e., $ \E_{c} [M_{c}]= \E_{c'} [M'_{c'}] = M'$, where $\E_{\star}[\cdot]$ denotes expectation over the  argument  in the subscript). These approximations together with Eq.~\ref{eq:DefSumMdash}   define an upper bound for the number of auxiliary bases, 
\EQ
\label{eq:SI_Ninequality}
N' \leq  \frac{a_0}{\langle\A\rangle } \left(1 - N \frac{\langle \A \rangle  }{a_0}\right).\EE

Using these relationships, we can now evaluated the variance of the affinities across random patterns $\chi$, as, 
\begin{align}
\label{eq:varRand0}
\var_\rand &= \var\left[ \A_0\sum_{\psi^\alpha\in\M} m^\alpha |\braket{\psi^\alpha|\chi}|^\Theta- \A_\rand\right]_\chi \\
\label{eq:varRand1} &= \A_0^2\, \var\left[  \sum_{c=1}^N M_c\,  \left| {\braket{\Psi^c(t)|\chi}}\right|^{\Theta}  +
   \sum_{c'=1}^{N'} M'_{c'}\,  \left| {\braket{\Phi^{c'}(t)|\chi}}\right|^{\Theta} \right]\\
\label{eq:varRand2}&=\A_0^2\, \sum_{c=1}^N \left(M_c\right)^2 \var\left(\left| \braket{\Psi^c |\chi}\right|^\Theta \right)  + \A_0^2\, \sum_{c'}^{N'} \left(M'_{c'}\right)^2 \var\left( \left| \braket{\Phi^{c'} |\chi}\right|^\Theta \right) \\
\label{eq:varRand3}
&=\A_0^2 \left(\sum_{c=1}^N \left(M_c\right)^2 + \sum_{c'}^{N'} \left(M'_{c'}\right)^2 \right) \var\left( \left| \braket{\psi|\chi}\right|^\Theta \right) \\
\label{eq:varRandfinal}
&\leq\A_0^2 \left( \frac{N}{a_0^2} \left(\var(\A) + \langle \A \rangle^2 \right)+  \frac{1}{a_0} \langle \A \rangle \left(1 - \frac{N}{a_0} \langle \A \rangle \right) \right) \times \frac{2^\Theta \left( \sqrt{\pi} \Gamma\left[\frac{1}{2}+ \Theta\right] - \Gamma	\left[\frac{1+\Theta}{2}\right]^2 \right) }{\pi L^\Theta} ,
\end{align}
where, we first used  Eq.~\ref{eq:DefArandSI2} to express the affinity function in Eq.~\ref{eq:varRand0} in terms of the bases in Eq.~\ref{eq:varRand1}. Given that the projections of a random  pattern along different bases are statistically independent from each other, we expressed the variance of the sum of these contributions in Eq.~\ref{eq:varRand1} as the sum of their variances in Eq.~\ref{eq:varRand2}. Next we used the fact that the overlaps of a random pattern with all the bases (i.e., $\braket{\Psi^c |\chi}, \,\braket{\Phi^{c'} |\chi}, \,\forall c,c'$) are Gaussian random variates with mean zero and variance $L/4$. Thus, the variance of these overlaps to the power $\Theta$ are equal for all bases, which we simply expressed as $\var\left[   \left|\braket{\psi|\chi}\right|^\Theta\right]$ in Eq.~\ref{eq:varRand3}. We then evaluated the term  $\var\left[   \left|\braket{\psi|\chi}\right|^\Theta\right]$ from the underlying Gaussian distribution, and used the definition of the effective weight $M_c$ in Eq.~\ref{eq:defMc} and the inequality in Eq.~\ref{eq:SI_Ninequality} to substitute $\sum_{c=1}^N (M_c)^2 +\sum_{c'=1}^{N'} (M'_{c'})^2 $ and arrive at the upper bound of the variance for the affinity for random patterns in  Eq.~\ref{eq:varRandfinal}. The results in Eq.~\ref{eq:varRandfinal}  match very well with simulations for the affinity function with the shape parameter $\Theta=2$ (Fig~\ref{fig:STD_Rand_SI}). 

\section{Mapping between immune memory repertoires and the Hopfield model}
\label{sec:MapToHop}
The Hopfield network is among the most frequently used models for associative memory~\cite{Hopfield:1982fq}. A classical Hopfield network  describes a fully connected graph with interaction matrix $J_{ij}$. A binary pattern $\psi$ of length $L$ presented to the network $J$ is assigned an energy $E(\psi,J)$,
\begin{align}
\label{eq:Standrd_Hop_e}
E(\psi,J) = -\frac{1}{2L} \sum_{i,j}^L J_{i,j} \psi_i \psi_j.
\end{align}

Hopfiled networks can learn and store an associative memory of the presented patterns as the minima of the energy landscape~\cite{Hopfield:1982fq}. One way to construct such network is by Hebbian learning, whereby the network is updated with a learning rate $\lambda$ upon an encounter with a pattern $\psi(t)$ at time $t$, 
\begin{equation}
J_{i,j}(t+1) = 
\begin{cases}
(1-\lambda)\,J_{i,j} (t) +\lambda \,\psi_i (t)\psi_j (t),&\text{if }i\neq j;\\
0,&\text{otherwise.}
\end{cases}
\label{eq:learningrule_Hop_SI}
\end{equation} 

In the supplementary information of ref.~\cite{Schnaack:Hop1}, we show that this learning rule can be expressed as 
\begin{equation}
J(t+1) = (1-\lambda) J(t)  + \lambda \left( L \ket{\psi(t)}\bra{\psi(t)} - \mathds{1} \right).
\label{eq:learningrule_Hop_braket_SI}
\end{equation} 
Using this expression for the network $J(T)$ at time $T$ with an encounter history with patterns $\psi(t)$ (for $t\leq T$), we can evaluate the energy $E(\chi,J(T))$ of an arbitrary pattern $\chi$  presented to the network as,
\begin{align}
\nonumber
E(\chi,J(T)) &=-\frac{1}{2} \braket{\chi|J(T)|\chi} \\
\nonumber &= -\frac{L}{2}\sum_{t=1}^{T-1}\lambda (1-\lambda)^{T - 1 -t} \braket{\chi|\psi(t)}\braket{\psi(t)|\chi} + \frac{1}{2}\sum_{t=1}^{T-1}\lambda (1-\lambda)^{T - 1 -t} \braket{\chi|\mathds{1}|\chi} \\
 \nonumber &{=} -\frac{L}{2}\sum_{t=1}^{T-1}\lambda (1-\lambda)^{T - 1 -t}\, |\braket{\psi(t)|\chi}|^2 +\frac{1}{2} +\mathcal O\left((1-\lambda)^T \right) \\
 \label{eq:Hop_energy_braket}
&{=} -\frac{L}{2}\sum_{t=1}^{T-1}m_t\, |\braket{\psi(t)|\chi}|^2 +\frac{1}{2} + \mathcal O\left((1-\lambda)^T\right) 
\end{align}
with the time-dependent weights, $m_t = \lambda (1-\lambda)^{T - 1 -t}$. Note that the correction terms vanish in the limit of large time $T$ and that the weights  sum up to one, i.e., $\sum m_t =1$. In this limit, the energy function $E(\chi,J)$  corresponds to the  affinity function given in Eq.~\ref{eq:affinity} of the main text, for the choice of the shape parameter $\Theta=2$, the energy scale $\A_0 = -L/2$, and the random energy $\A_\rand = -1/2$.

\section{Numerical methods}
\label{sec:Nummet}
As discussed in Section~\ref{sec:MapToHop}, for the shape parameter   $\Theta=2$  the affinity function in Eq.~\ref{eq:affinity} maps onto the energy function of a Hopfield network with Hebbian learning~\cite{Hopfield:1982fq,workingMem87,Schnaack:Hop1}, which  is easily tractable with numerical techniques. Specifically,  using a Hopfield model for simulating the repertoire problem with $\Theta=2$  has the advantage that we only need to keep track of the interaction matrix $J_{i,j}$ of size $L^2$ as opposed to all  the $2^L$ memory wights $m^\alpha$. This dramatic reduction in complexity makes the numerical simulations for $\Theta=2$ highly efficient. 

For the simulations, we use the same approach as in \cite{Schnaack:Hop1}. We  initialize the interaction matrix $J$ of size $L \times L$ with all entries set to  zero  $J_{i,j}=0$ and choose $N$ independent random patterns of size $L$.

Before collecting any data, we first update the network until it  reaches a quasi-stationary state so  that it no longer depends on the initial condition. Since the stored memory of a pattern  within the  network decays  as $(1-\lambda)^s$ with the number the number update steps  $s$ since the original encounter, we  update the network  following the initialization by $s_{\rm stat.} = \operatorname{ceil} \left(  \frac{\log10^{-5}}{\log(1-\lambda)} \right)$  steps for the network to reach a quasi-stationary state. {This criterion ensures that $(1-\lambda)^{s_\mathrm{stat.} }\leq 10^{-5}$ and the memory of the initial state is removed. Moreover, this criterion implies that in the  quasi-stationary state the memory weights become normalized i.e., $1- \sum_\alpha m^\alpha(s_\mathrm{stat.}) <10^{-5}$, after starting from a no memory initial state of $\sum_\alpha m^\alpha(0) =0$. During each step of this preparation phase, we also evolve the patterns, whereby we flip the spins of patterns within each class   at  rate $\mu$. We then randomly choose one of the patterns to present to the network and update the network according to the learning rule in  Eq.~\ref{eq:learningrule_Hop_SI}.

After reaching the quasi-stationary state of memory, we collect data from the network over $10^4$ steps. During this process, the patterns evolve with rate $\mu$ and are presented at a random order to the network. We record the affinity of each presented pattern before updating the network. At each step we also record the affinity of a randomly generated pattern that belongs to none of the previously encountered  pattern classes. For each learning rate $\lambda$, we repeat this process for $50$ independent initial sets of pattern classes $\{\Psi^c\}$. Overall, we perform  a total of  $5 \times 10^5$ affinity measurements on both the  previously encountered and the random patterns.

\newpage

\begin{figure}[ht!]
\centering
\includegraphics[]{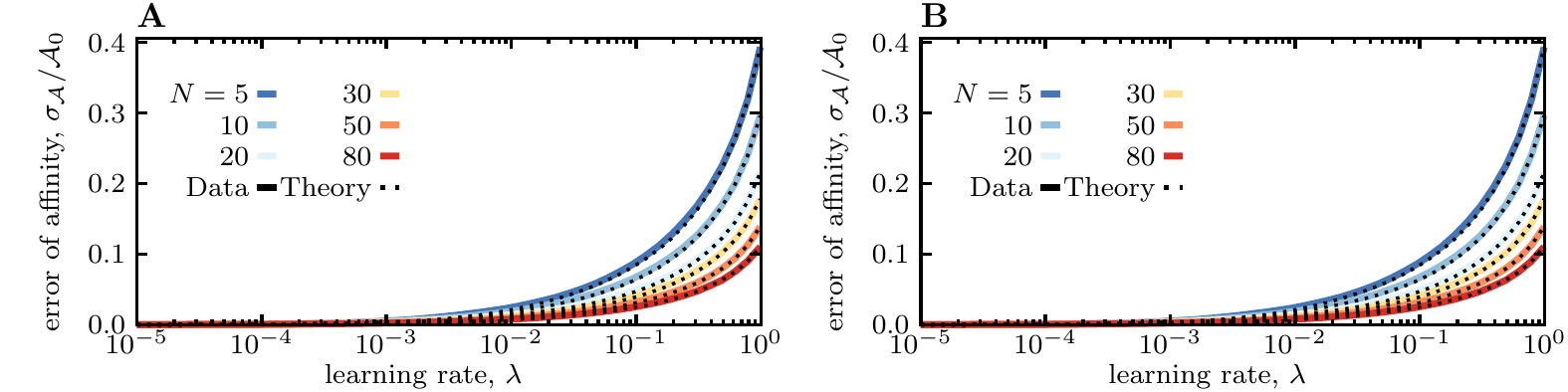}
 \caption{{\bf Standard deviation of affinities for pre-encountered patterns}. Solid lines shown the numerical estimates for  the standard deviation  of pattern affinities ($\sigma_\A$) divided by the scale of the affinity ($\A_0$) for different numbers of patterns $N$  (colors), using Hopfield model to simulate of repertoires. Dotted lines show the analytical estimates, using {\bf (A)} the full solution in Eq.~\ref{eq:theory_variance_energy}, and  {\bf (B)}  the approximation in Eqs.~\ref{eq:cumulants}~\&~\ref{eq:cumulantsSI}. Simulation parameters: $L=100$, $\mu_\eff = 0.01$, $\Theta =2$.}
\label{fig:STD_patt_SI} 
\end{figure}

\vspace{3cm}

\begin{figure}[ht!]
\centering
\includegraphics[width=\textwidth]{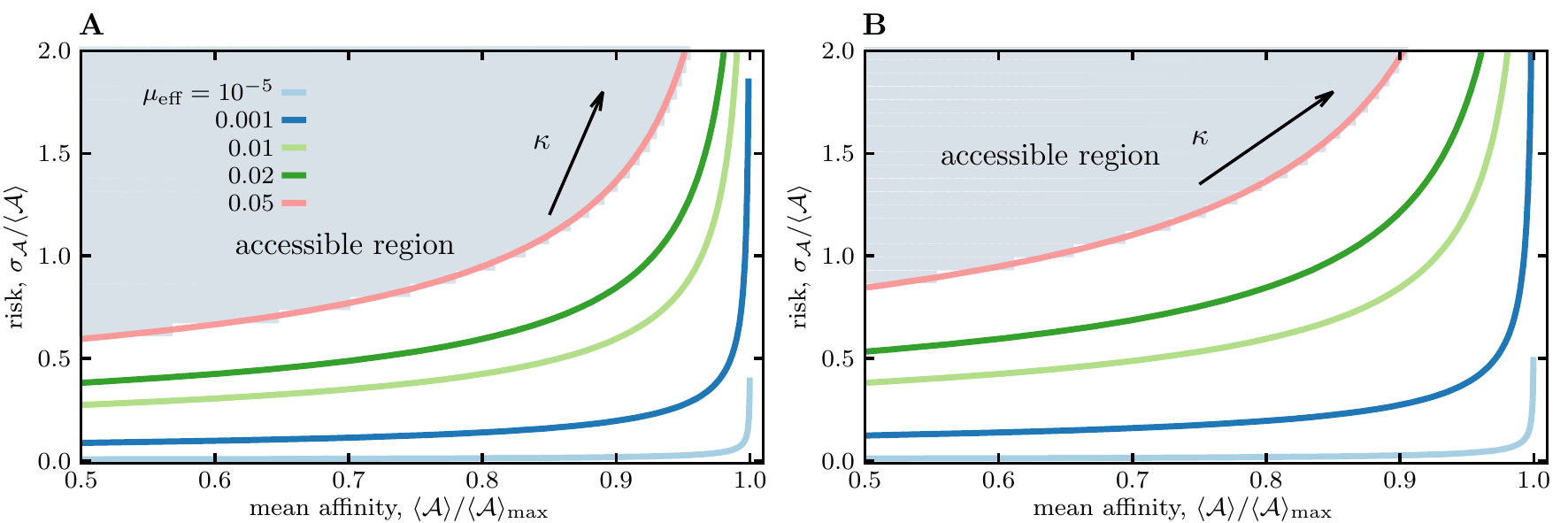}
 \caption{ {\bf Pareto front for risk-utility tradeoff of memory for evolving patterns.} Similar to Fig.~\ref{fig:Fig2} in the main text,  the risk-affinity Pareto front for the optimized objective function $\Q(\lambda^*)$ (Eq.~\ref{eq:Objective}) is shown for different mutation rates (colors) by varying the risk tolerance $\kappa$ along each line, for the shape parameters {\bf (A)} $\Theta =4$, and {\bf (B)} $\Theta =8$. In both cases, $N=200$.}
\label{fig:Pareto_SI} 
\end{figure}

\begin{figure}[ht!]
\centering
\includegraphics[]{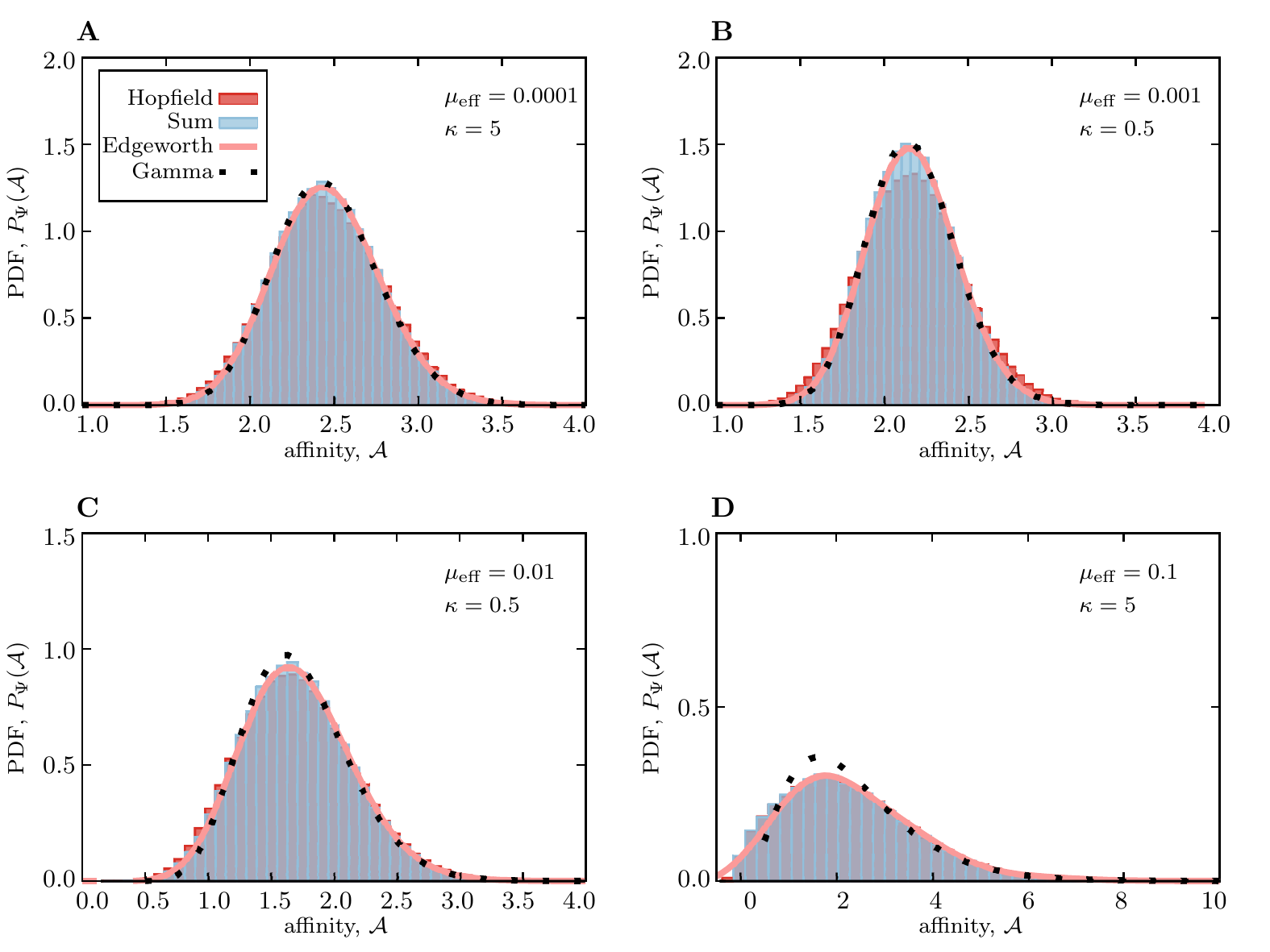}
 \caption{\textbf{Distribution of pattern affinities.} 
 The distribution of affinities between a memory repertoire and patterns from previously enchanted classes are shown  for different effective mutation rates $\mu_\eff$ and risk tolerance $\kappa$ in different panels. The distributions are characterized based on the simulations using Hopfield network (red; see  Appendix~\ref{sec:Nummet}), the process for deriving the cumulant-generating function in Eq~\ref{eq:mean_A_second_version}. (blue), the  Edgeworth approximation for the probability density function in Eq.~\ref{eq:Edgeworth} (pink), and the  Gamma distribution with the matching mean and variance (dotted lines). For small $(\mu_\eff \cdot \kappa)$ in ({\bf A-C}) all distributions are comparable and as suggested by the Kullback-Leibler divergence in  Eq.~\ref{eq:SI_DKL_values}, the  Gamma distribution is a good approximation to the underlying distribution of affinities.
For larger $(\mu_\eff \cdot \kappa)$ in  {\bf (D)},  the optimal learning rate becomes large (Eq.~\ref{eq:opt_lambda}), and the repertoire only remembers the most recently encountered patterns, resulting in the break down of the analytical approximations. Simulation parameters: $L=200$, $N=40$, and $\Theta =2$. }
\label{figHistogram_SI} 
\end{figure}

\begin{figure}[ht!]
\centering
\includegraphics[]{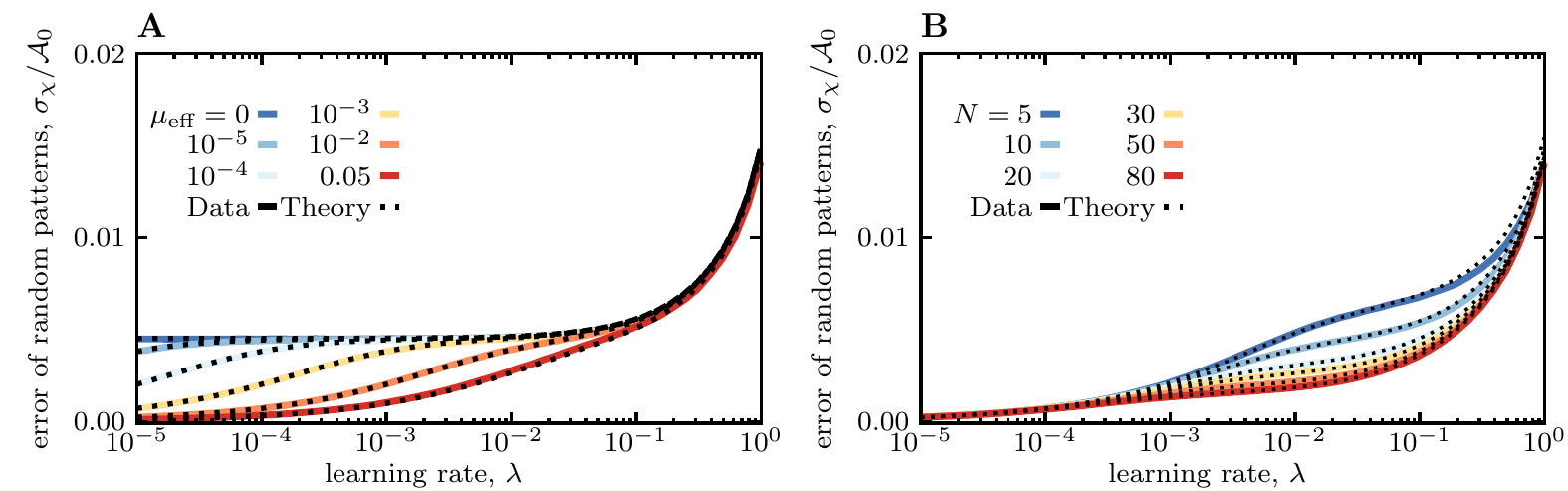}
 \caption{{\bf Standard deviation of affinities for random patterns}. 
 Solid lines show simulation results for $\Theta=2$ and dashed lines give the analytic result from Eq.~\ref{eq:varRandfinal}. \textbf{(A)} shows results for constant number of pattern classes $N = 30$ and for different effective mutation rates $\mu_\eff$  (colors). \textbf{(B)} shows results for constant $\mu_\eff = 0.01$ and for different numbers of pattern classes  $N$  (colors). In both cases we observe, that when the learning rate $\lambda$ is large and only one pattern class is stored in the system the fluctuations of random affinities are the largest as there is no self averaging  over all stored patterns. When the system can learn close to the optimum  (maximal affinity for stored patterns with small fluctuations) the prefactor in Eq.~\ref{eq:varRandfinal} reaches the na\"ive expectation $1/N$ and the fluctuations of random overlaps are averaged over all stored patterns. When the learning rate become small and the system can no longer follow the evolution of the patterns, the system effectively stores many additional states and the random fluctuations are averaged over all these states. As a consequence, the fluctuations $\sigma_\chi$ go to zero. Simulation parameters: $L=100$, $\Theta =2$. }
\label{fig:STD_Rand_SI} 
\end{figure}

\begin{figure}[ht!]
\centering
\includegraphics[]{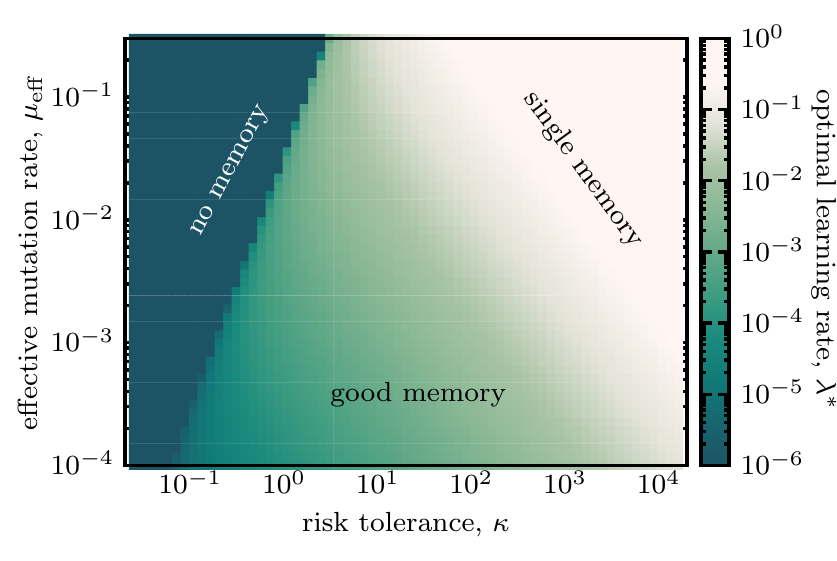}
 \caption{{\bf Optimal learning rate for three phases of memory.} The optimal learning rate $\lambda^*$ corresponding to the different  strategies   in Fig.~\ref{fig:Fig3} is shown for different values of   risk tolerance and mutation rate. Fast learning with $\lambda \sim 1$ corresponds to the phase of single memory storage (light), where only the memory of most recent encounter is retained. On the other hand,  slow learning  $\lambda \ll 1$ corresponds to the phase where effectively no memory is stored (dark). The triangle of good memory is associated with intermediate rates of learning. Simulation parameters: $L=200$, $N=40$, and $\Theta=2$. }
\label{fig:Optimal_learning_heat_SI} 
\end{figure}

\begin{figure}[ht!]
\centering
\includegraphics[]{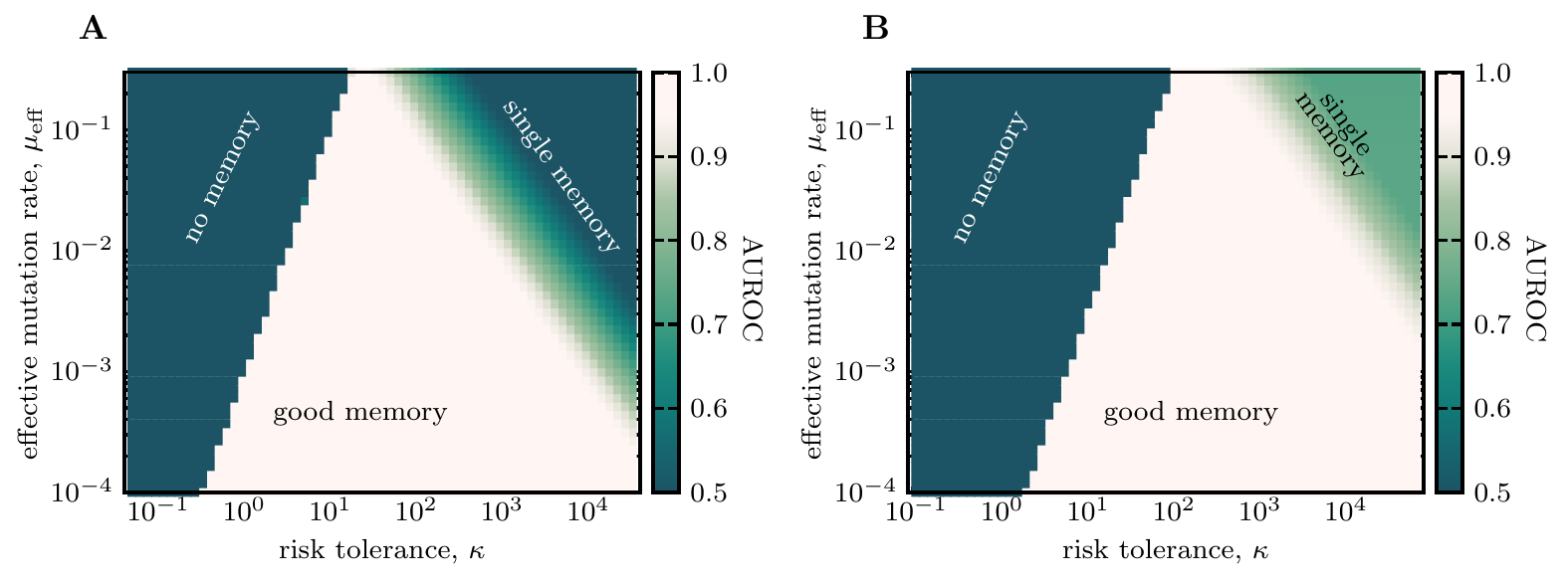}
 \caption{\textbf{Three phases of memory discrimination for different shape parameters of the affinity function.}  
 Similar to Fig.~\ref{fig:Fig3} the phase diagram shows the discrimination ability of repertoires (AUROC) between familiar patterns with prior encounter history  and random patterns, for different shape parameters of the affinity function {\bf (A)} $\Theta =4$ , and {\bf (B)} $\Theta =8$. Other parameters: Parameters: $L=200$, $N=40$.}
\label{fig:ROC_SI} 
\end{figure}

\begin{figure}[ht!]
\centering
\includegraphics[]{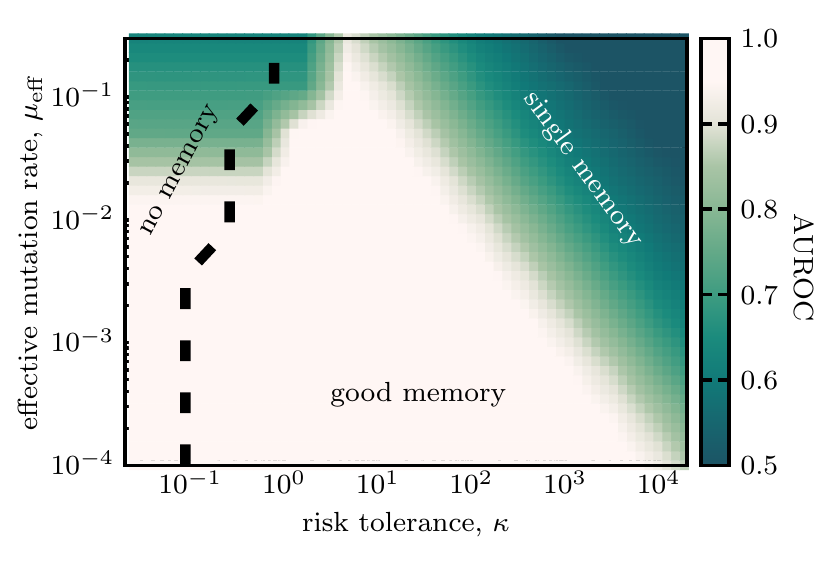}
 \caption{\textbf{Three phases of memory discrimination uncovered with simulations of Hopfield networks.} Similar to Fig.~\ref{fig:Fig3} the phase diagram shows discrimination ability of repertoires (AUROC) between familiar patterns with prior encounter history  and random patterns. The phase diagram is acquired by direct simulation of memory, using the correspondence between repertoires with  shape parameter $\Theta =2$ and Hopfield network; see Appendix~\ref{sec:Nummet} for numeral technique. The low risk region on the left side of the dashed line is not accessible by simulations, which explains the differences between the simulations and results of the analytic approximation shown in Fig.~\ref{fig:Fig3}.  Parameters: $L=200$, $N=40$, and $\Theta =2$.}
\label{fig:ROC_DATA_SI} 
\end{figure}


\begin{thebibliography}{30}%
\makeatletter
\providecommand \@ifxundefined [1]{%
 \@ifx{#1\undefined}
}%
\providecommand \@ifnum [1]{%
 \ifnum #1\expandafter \@firstoftwo
 \else \expandafter \@secondoftwo
 \fi
}%
\providecommand \@ifx [1]{%
 \ifx #1\expandafter \@firstoftwo
 \else \expandafter \@secondoftwo
 \fi
}%
\providecommand \natexlab [1]{#1}%
\providecommand \enquote  [1]{``#1''}%
\providecommand \bibnamefont  [1]{#1}%
\providecommand \bibfnamefont [1]{#1}%
\providecommand \citenamefont [1]{#1}%
\providecommand \href@noop [0]{\@secondoftwo}%
\providecommand \href [0]{\begingroup \@sanitize@url \@href}%
\providecommand \@href[1]{\@@startlink{#1}\@@href}%
\providecommand \@@href[1]{\endgroup#1\@@endlink}%
\providecommand \@sanitize@url [0]{\catcode `\\12\catcode `\$12\catcode
  `\&12\catcode `\#12\catcode `\^12\catcode `\_12\catcode `\%12\relax}%
\providecommand \@@startlink[1]{}%
\providecommand \@@endlink[0]{}%
\providecommand \url  [0]{\begingroup\@sanitize@url \@url }%
\providecommand \@url [1]{\endgroup\@href {#1}{\urlprefix }}%
\providecommand \urlprefix  [0]{URL }%
\providecommand \Eprint [0]{\href }%
\providecommand \doibase [0]{https://doi.org/}%
\providecommand \selectlanguage [0]{\@gobble}%
\providecommand \bibinfo  [0]{\@secondoftwo}%
\providecommand \bibfield  [0]{\@secondoftwo}%
\providecommand \translation [1]{[#1]}%
\providecommand \BibitemOpen [0]{}%
\providecommand \bibitemStop [0]{}%
\providecommand \bibitemNoStop [0]{.\EOS\space}%
\providecommand \EOS [0]{\spacefactor3000\relax}%
\providecommand \BibitemShut  [1]{\csname bibitem#1\endcsname}%
\let\auto@bib@innerbib\@empty
\bibitem [{\citenamefont {Goodfellow}\ \emph {et~al.}(2016)\citenamefont
  {Goodfellow}, \citenamefont {Bengio},\ and\ \citenamefont
  {Courville}}]{Goodfellow-et-al-2016}%
  \BibitemOpen
  \bibfield  {author} {\bibinfo {author} {\bibfnamefont {I.}~\bibnamefont
  {Goodfellow}}, \bibinfo {author} {\bibfnamefont {Y.}~\bibnamefont {Bengio}},\
  and\ \bibinfo {author} {\bibfnamefont {A.}~\bibnamefont {Courville}},\
  }\href@noop {} {\emph {\bibinfo {title} {Deep Learning}}}\ (\bibinfo
  {publisher} {MIT Press},\ \bibinfo {year} {2016})\ \bibinfo {note}
  {\url{http://www.deeplearningbook.org}}\BibitemShut {NoStop}%
\bibitem [{\citenamefont {Soltoggio}\ \emph {et~al.}(2018)\citenamefont
  {Soltoggio}, \citenamefont {Stanley},\ and\ \citenamefont
  {Risi}}]{soltoggio_born_2018}%
  \BibitemOpen
  \bibfield  {author} {\bibinfo {author} {\bibfnamefont {A.}~\bibnamefont
  {Soltoggio}}, \bibinfo {author} {\bibfnamefont {K.~O.}\ \bibnamefont
  {Stanley}},\ and\ \bibinfo {author} {\bibfnamefont {S.}~\bibnamefont
  {Risi}},\ }\bibfield  {title} {\bibinfo {title} {Born to learn: {The}
  inspiration, progress, and future of evolved plastic artificial neural
  networks},\ }\href {https://doi.org/10.1016/j.neunet.2018.07.013} {\bibfield
  {journal} {\bibinfo  {journal} {Neural Networks}\ }\textbf {\bibinfo {volume}
  {108}},\ \bibinfo {pages} {48} (\bibinfo {year} {2018})}\BibitemShut
  {NoStop}%
\bibitem [{\citenamefont {Mehta}\ \emph {et~al.}(2019)\citenamefont {Mehta},
  \citenamefont {Bukov}, \citenamefont {Wang}, \citenamefont {Day},
  \citenamefont {Richardson}, \citenamefont {Fisher},\ and\ \citenamefont
  {Schwab}}]{Mehta:2019pr}%
  \BibitemOpen
  \bibfield  {author} {\bibinfo {author} {\bibfnamefont {P.}~\bibnamefont
  {Mehta}}, \bibinfo {author} {\bibfnamefont {M.}~\bibnamefont {Bukov}},
  \bibinfo {author} {\bibfnamefont {C.-H.}\ \bibnamefont {Wang}}, \bibinfo
  {author} {\bibfnamefont {A.~G.~R.}\ \bibnamefont {Day}}, \bibinfo {author}
  {\bibfnamefont {C.}~\bibnamefont {Richardson}}, \bibinfo {author}
  {\bibfnamefont {C.~K.}\ \bibnamefont {Fisher}},\ and\ \bibinfo {author}
  {\bibfnamefont {D.~J.}\ \bibnamefont {Schwab}},\ }\bibfield  {title}
  {\bibinfo {title} {A high-bias, low-variance introduction to {Machine}
  {Learning} for physicists},\ }\href
  {https://doi.org/10.1016/j.physrep.2019.03.001} {\bibfield  {journal}
  {\bibinfo  {journal} {Physics Reports}\ }\textbf {\bibinfo {volume} {810}},\
  \bibinfo {pages} {1} (\bibinfo {year} {2019})},\ \bibinfo {note} {arXiv:
  1803.08823}\BibitemShut {NoStop}%
\bibitem [{\citenamefont {Janeway}\ \emph {et~al.}(2001)\citenamefont
  {Janeway}, \citenamefont {Travers}, \citenamefont {Walport},\ and\
  \citenamefont {Schlomchik}}]{Janeway:2001te}%
  \BibitemOpen
  \bibfield  {author} {\bibinfo {author} {\bibfnamefont {C.}~\bibnamefont
  {Janeway}}, \bibinfo {author} {\bibfnamefont {P.}~\bibnamefont {Travers}},
  \bibinfo {author} {\bibfnamefont {M.}~\bibnamefont {Walport}},\ and\ \bibinfo
  {author} {\bibfnamefont {M.}~\bibnamefont {Schlomchik}},\ }\href@noop {}
  {\emph {\bibinfo {title} {{Immunobiology}}}},\ \bibinfo {edition} {5th}\
  ed.,\ The Immune System in Health and Disease\ (\bibinfo  {publisher}
  {Garland Science},\ \bibinfo {address} {New York},\ \bibinfo {year}
  {2001})\BibitemShut {NoStop}%
\bibitem [{\citenamefont {Barrangou}\ and\ \citenamefont
  {Marraffini}(2014)}]{Barrangou:2014ht}%
  \BibitemOpen
  \bibfield  {author} {\bibinfo {author} {\bibfnamefont {R.}~\bibnamefont
  {Barrangou}}\ and\ \bibinfo {author} {\bibfnamefont {L.~A.}\ \bibnamefont
  {Marraffini}},\ }\bibfield  {title} {\bibinfo {title} {{CRISPR-Cas systems:
  Prokaryotes upgrade to adaptive immunity}},\ }\href@noop {} {\bibfield
  {journal} {\bibinfo  {journal} {Molecular cell}\ }\textbf {\bibinfo {volume}
  {54}},\ \bibinfo {pages} {234} (\bibinfo {year} {2014})}\BibitemShut
  {NoStop}%
\bibitem [{\citenamefont {Altan-Bonnet}\ \emph {et~al.}(2020)\citenamefont
  {Altan-Bonnet}, \citenamefont {Mora},\ and\ \citenamefont
  {Walczak}}]{AltanBonnet:2020hk}%
  \BibitemOpen
  \bibfield  {author} {\bibinfo {author} {\bibfnamefont {G.}~\bibnamefont
  {Altan-Bonnet}}, \bibinfo {author} {\bibfnamefont {T.}~\bibnamefont {Mora}},\
  and\ \bibinfo {author} {\bibfnamefont {A.~M.}\ \bibnamefont {Walczak}},\
  }\bibfield  {title} {\bibinfo {title} {{Quantitative immunology for
  physicists}},\ }\href@noop {} {\bibfield  {journal} {\bibinfo  {journal}
  {Physics Reports}\ }\textbf {\bibinfo {volume} {849}},\ \bibinfo {pages} {1}
  (\bibinfo {year} {2020})}\BibitemShut {NoStop}%
\bibitem [{\citenamefont {Bradde}\ \emph {et~al.}(2020)\citenamefont {Bradde},
  \citenamefont {Nourmohammad}, \citenamefont {Goyal},\ and\ \citenamefont
  {Balasubramanian}}]{Bradde:2020kb}%
  \BibitemOpen
  \bibfield  {author} {\bibinfo {author} {\bibfnamefont {S.}~\bibnamefont
  {Bradde}}, \bibinfo {author} {\bibfnamefont {A.}~\bibnamefont
  {Nourmohammad}}, \bibinfo {author} {\bibfnamefont {S.}~\bibnamefont
  {Goyal}},\ and\ \bibinfo {author} {\bibfnamefont {V.}~\bibnamefont
  {Balasubramanian}},\ }\bibfield  {title} {\bibinfo {title} {{The size of the
  immune repertoire of bacteria}},\ }\href@noop {} {\bibfield  {journal}
  {\bibinfo  {journal} {Proc. Natl. Acad. Sci. U.S.A.}\ }\textbf {\bibinfo
  {volume} {117}},\ \bibinfo {pages} {5144} (\bibinfo {year}
  {2020})}\BibitemShut {NoStop}%
\bibitem [{\citenamefont {Schnaack}\ and\ \citenamefont
  {Nourmohammad}(2021)}]{Schnaack:2020vb}%
  \BibitemOpen
  \bibfield  {author} {\bibinfo {author} {\bibfnamefont {O.~H.}\ \bibnamefont
  {Schnaack}}\ and\ \bibinfo {author} {\bibfnamefont {A.}~\bibnamefont
  {Nourmohammad}},\ }\bibfield  {title} {\bibinfo {title} {Optimal evolutionary
  decision-making to store immune memory},\ }\href
  {https://doi.org/10.7554/eLife.61346} {\bibfield  {journal} {\bibinfo
  {journal} {eLife}\ }\textbf {\bibinfo {volume} {10}},\ \bibinfo {pages}
  {e61346} (\bibinfo {year} {2021})}\BibitemShut {NoStop}%
\bibitem [{\citenamefont {Schnaack}\ \emph {et~al.}(2021)\citenamefont
  {Schnaack}, \citenamefont {Peliti},\ and\ \citenamefont
  {Nourmohammad}}]{Schnaack:Hop1}%
  \BibitemOpen
  \bibfield  {author} {\bibinfo {author} {\bibfnamefont {O.~H.}\ \bibnamefont
  {Schnaack}}, \bibinfo {author} {\bibfnamefont {L.}~\bibnamefont {Peliti}},\
  and\ \bibinfo {author} {\bibfnamefont {A.}~\bibnamefont {Nourmohammad}},\
  }\bibfield  {title} {\bibinfo {title} {Learning and organization of memory
  for evolving patterns},\ }\href@noop {} {\bibfield  {journal} {\bibinfo
  {journal} {arXiv:2106.02186 [physics]}\ } (\bibinfo {year} {2021})},\
  \bibinfo {note} {arXiv: 2106.02186}\BibitemShut {NoStop}%
\bibitem [{\citenamefont {Hopfield}(1982)}]{Hopfield:1982fq}%
  \BibitemOpen
  \bibfield  {author} {\bibinfo {author} {\bibfnamefont {J.~J.}\ \bibnamefont
  {Hopfield}},\ }\bibfield  {title} {\bibinfo {title} {{Neural networks and
  physical systems with emergent collective computational abilities}},\
  }\href@noop {} {\bibfield  {journal} {\bibinfo  {journal} {Proc. Natl. Acad.
  Sci. U.S.A.}\ }\textbf {\bibinfo {volume} {79}},\ \bibinfo {pages} {2554}
  (\bibinfo {year} {1982})}\BibitemShut {NoStop}%
\bibitem [{\citenamefont {Hebb}(1949)}]{Hebb:1949vs}%
  \BibitemOpen
  \bibfield  {author} {\bibinfo {author} {\bibfnamefont {D.~O.}\ \bibnamefont
  {Hebb}},\ }\href@noop {} {\emph {\bibinfo {title} {{The Organization of
  Behavior: A Neuropsychological Theory}}}}\ (\bibinfo  {publisher} {Wiley},\
  \bibinfo {address} {New York},\ \bibinfo {year} {1949})\BibitemShut {NoStop}%
\bibitem [{\citenamefont {Mezard}\ \emph {et~al.}(1986)\citenamefont {Mezard},
  \citenamefont {Nadal},\ and\ \citenamefont {Toulouse}}]{workingMem87}%
  \BibitemOpen
  \bibfield  {author} {\bibinfo {author} {\bibfnamefont {M.}~\bibnamefont
  {Mezard}}, \bibinfo {author} {\bibfnamefont {J.~P.}\ \bibnamefont {Nadal}},\
  and\ \bibinfo {author} {\bibfnamefont {G.}~\bibnamefont {Toulouse}},\
  }\bibfield  {title} {\bibinfo {title} {Solvable models of working memories},\
  }\href@noop {} {\bibfield  {journal} {\bibinfo  {journal} {J. Physique}\
  }\textbf {\bibinfo {volume} {47}},\ \bibinfo {pages} {1457} (\bibinfo {year}
  {1986})}\BibitemShut {NoStop}%
\bibitem [{\citenamefont {Amit}\ \emph {et~al.}(1985)\citenamefont {Amit},
  \citenamefont {Gutfreund},\ and\ \citenamefont {Sompolinsky}}]{Amit:1985bo}%
  \BibitemOpen
  \bibfield  {author} {\bibinfo {author} {\bibfnamefont {D.~J.}\ \bibnamefont
  {Amit}}, \bibinfo {author} {\bibfnamefont {H.}~\bibnamefont {Gutfreund}},\
  and\ \bibinfo {author} {\bibfnamefont {H.}~\bibnamefont {Sompolinsky}},\
  }\bibfield  {title} {\bibinfo {title} {Storing infinite numbers of patterns
  in a spin-glass model of neural networks},\ }\href@noop {} {\bibfield
  {journal} {\bibinfo  {journal} {Phys. Rev. Lett.}\ }\textbf {\bibinfo
  {volume} {55}},\ \bibinfo {pages} {1530} (\bibinfo {year}
  {1985})}\BibitemShut {NoStop}%
\bibitem [{\citenamefont {Steuer}(1986)}]{steuer_multiple_1986}%
  \BibitemOpen
  \bibfield  {author} {\bibinfo {author} {\bibfnamefont {R.~E.}\ \bibnamefont
  {Steuer}},\ }\href@noop {} {\emph {\bibinfo {title} {Multiple criteria
  optimization: theory, computation, and application}}},\ Wiley series in
  probability and mathematical statistics\ (\bibinfo  {publisher} {Wiley},\
  \bibinfo {address} {New York},\ \bibinfo {year} {1986})\BibitemShut {NoStop}%
\bibitem [{\citenamefont {Sen}(1993)}]{sen_markets_1993}%
  \BibitemOpen
  \bibfield  {author} {\bibinfo {author} {\bibfnamefont {A.}~\bibnamefont
  {Sen}},\ }\bibfield  {title} {\bibinfo {title} {Markets and {Freedoms}:
  {Achievements} and {Limitations} of the {Market} {Mechanism} in {Promoting}
  {Individual} {Freedoms}},\ }\href@noop {} {\bibfield  {journal} {\bibinfo
  {journal} {Oxf. Econ. Pap.}\ }\textbf {\bibinfo {volume} {45}},\ \bibinfo
  {pages} {519} (\bibinfo {year} {1993})}\BibitemShut {NoStop}%
\bibitem [{\citenamefont {Shoval}\ \emph {et~al.}(2012)\citenamefont {Shoval},
  \citenamefont {Sheftel}, \citenamefont {Shinar}, \citenamefont {Hart},
  \citenamefont {Ramote}, \citenamefont {Mayo}, \citenamefont {Dekel},
  \citenamefont {Kavanagh},\ and\ \citenamefont
  {Alon}}]{shoval_evolutionary_2012}%
  \BibitemOpen
  \bibfield  {author} {\bibinfo {author} {\bibfnamefont {O.}~\bibnamefont
  {Shoval}}, \bibinfo {author} {\bibfnamefont {H.}~\bibnamefont {Sheftel}},
  \bibinfo {author} {\bibfnamefont {G.}~\bibnamefont {Shinar}}, \bibinfo
  {author} {\bibfnamefont {Y.}~\bibnamefont {Hart}}, \bibinfo {author}
  {\bibfnamefont {O.}~\bibnamefont {Ramote}}, \bibinfo {author} {\bibfnamefont
  {A.}~\bibnamefont {Mayo}}, \bibinfo {author} {\bibfnamefont {E.}~\bibnamefont
  {Dekel}}, \bibinfo {author} {\bibfnamefont {K.}~\bibnamefont {Kavanagh}},\
  and\ \bibinfo {author} {\bibfnamefont {U.}~\bibnamefont {Alon}},\ }\bibfield
  {title} {\bibinfo {title} {Evolutionary {Trade}-{Offs}, {Pareto}
  {Optimality}, and the {Geometry} of {Phenotype} {Space}},\ }\href
  {https://doi.org/10.1126/science.1217405} {\bibfield  {journal} {\bibinfo
  {journal} {Science}\ }\textbf {\bibinfo {volume} {336}},\ \bibinfo {pages}
  {1157} (\bibinfo {year} {2012})}\BibitemShut {NoStop}%
\bibitem [{\citenamefont {Schuetz}\ \emph {et~al.}(2012)\citenamefont
  {Schuetz}, \citenamefont {Zamboni}, \citenamefont {Zampieri}, \citenamefont
  {Heinemann},\ and\ \citenamefont {Sauer}}]{schuetz_multidimensional_2012}%
  \BibitemOpen
  \bibfield  {author} {\bibinfo {author} {\bibfnamefont {R.}~\bibnamefont
  {Schuetz}}, \bibinfo {author} {\bibfnamefont {N.}~\bibnamefont {Zamboni}},
  \bibinfo {author} {\bibfnamefont {M.}~\bibnamefont {Zampieri}}, \bibinfo
  {author} {\bibfnamefont {M.}~\bibnamefont {Heinemann}},\ and\ \bibinfo
  {author} {\bibfnamefont {U.}~\bibnamefont {Sauer}},\ }\bibfield  {title}
  {\bibinfo {title} {Multidimensional {Optimality} of {Microbial}
  {Metabolism}},\ }\href {https://doi.org/10.1126/science.1216882} {\bibfield
  {journal} {\bibinfo  {journal} {Science}\ }\textbf {\bibinfo {volume}
  {336}},\ \bibinfo {pages} {601} (\bibinfo {year} {2012})}\BibitemShut
  {NoStop}%
\bibitem [{\citenamefont {Hart}\ \emph {et~al.}(2015)\citenamefont {Hart},
  \citenamefont {Sheftel}, \citenamefont {Hausser}, \citenamefont {Szekely},
  \citenamefont {Ben-Moshe}, \citenamefont {Korem}, \citenamefont {Tendler},
  \citenamefont {Mayo},\ and\ \citenamefont {Alon}}]{hart_inferring_2015}%
  \BibitemOpen
  \bibfield  {author} {\bibinfo {author} {\bibfnamefont {Y.}~\bibnamefont
  {Hart}}, \bibinfo {author} {\bibfnamefont {H.}~\bibnamefont {Sheftel}},
  \bibinfo {author} {\bibfnamefont {J.}~\bibnamefont {Hausser}}, \bibinfo
  {author} {\bibfnamefont {P.}~\bibnamefont {Szekely}}, \bibinfo {author}
  {\bibfnamefont {N.~B.}\ \bibnamefont {Ben-Moshe}}, \bibinfo {author}
  {\bibfnamefont {Y.}~\bibnamefont {Korem}}, \bibinfo {author} {\bibfnamefont
  {A.}~\bibnamefont {Tendler}}, \bibinfo {author} {\bibfnamefont {A.~E.}\
  \bibnamefont {Mayo}},\ and\ \bibinfo {author} {\bibfnamefont
  {U.}~\bibnamefont {Alon}},\ }\bibfield  {title} {\bibinfo {title} {Inferring
  biological tasks using {Pareto} analysis of high-dimensional data},\ }\href
  {https://doi.org/10.1038/nmeth.3254} {\bibfield  {journal} {\bibinfo
  {journal} {Nat Methods}\ }\textbf {\bibinfo {volume} {12}},\ \bibinfo {pages}
  {233} (\bibinfo {year} {2015})}\BibitemShut {NoStop}%
\bibitem [{\citenamefont {Szekely}\ \emph {et~al.}(2015)\citenamefont
  {Szekely}, \citenamefont {Korem}, \citenamefont {Moran}, \citenamefont
  {Mayo},\ and\ \citenamefont {Alon}}]{szekely_mass-longevity_2015}%
  \BibitemOpen
  \bibfield  {author} {\bibinfo {author} {\bibfnamefont {P.}~\bibnamefont
  {Szekely}}, \bibinfo {author} {\bibfnamefont {Y.}~\bibnamefont {Korem}},
  \bibinfo {author} {\bibfnamefont {U.}~\bibnamefont {Moran}}, \bibinfo
  {author} {\bibfnamefont {A.}~\bibnamefont {Mayo}},\ and\ \bibinfo {author}
  {\bibfnamefont {U.}~\bibnamefont {Alon}},\ }\bibfield  {title} {\bibinfo
  {title} {The {Mass}-{Longevity} {Triangle}: {Pareto} {Optimality} and the
  {Geometry} of {Life}-{History} {Trait} {Space}},\ }\href
  {https://doi.org/10.1371/journal.pcbi.1004524} {\bibfield  {journal}
  {\bibinfo  {journal} {PLoS Comput Biol}\ }\textbf {\bibinfo {volume} {11}},\
  \bibinfo {pages} {e1004524} (\bibinfo {year} {2015})}\BibitemShut {NoStop}%
\bibitem [{\citenamefont {Seoane}\ and\ \citenamefont
  {Sol\'e}(2015)}]{seoane_phase_2015}%
  \BibitemOpen
  \bibfield  {author} {\bibinfo {author} {\bibfnamefont {L.~F.}\ \bibnamefont
  {Seoane}}\ and\ \bibinfo {author} {\bibfnamefont {R.}~\bibnamefont {Sol\'e}},\
  }\bibfield  {title} {\bibinfo {title} {Phase transitions in {Pareto} optimal
  complex networks},\ }\href {https://doi.org/10.1103/PhysRevE.92.032807}
  {\bibfield  {journal} {\bibinfo  {journal} {Phys. Rev. E}\ }\textbf {\bibinfo
  {volume} {92}},\ \bibinfo {pages} {032807} (\bibinfo {year}
  {2015})}\BibitemShut {NoStop}%
\bibitem [{\citenamefont {Tendler}\ \emph {et~al.}(2015)\citenamefont
  {Tendler}, \citenamefont {Mayo},\ and\ \citenamefont
  {Alon}}]{tendler_evolutionary_2015}%
  \BibitemOpen
  \bibfield  {author} {\bibinfo {author} {\bibfnamefont {A.}~\bibnamefont
  {Tendler}}, \bibinfo {author} {\bibfnamefont {A.}~\bibnamefont {Mayo}},\ and\
  \bibinfo {author} {\bibfnamefont {U.}~\bibnamefont {Alon}},\ }\bibfield
  {title} {\bibinfo {title} {Evolutionary tradeoffs, {Pareto} optimality and
  the morphology of ammonite shells},\ }\href
  {https://doi.org/10.1186/s12918-015-0149-z} {\bibfield  {journal} {\bibinfo
  {journal} {BMC Syst Biol}\ }\textbf {\bibinfo {volume} {9}},\ \bibinfo
  {pages} {12} (\bibinfo {year} {2015})}\BibitemShut {NoStop}%
\bibitem [{\citenamefont {Ko√ßillari}\ \emph {et~al.}(2018)\citenamefont
  {Ko\c illari}, \citenamefont {Fariselli}, \citenamefont {Trovato},
  \citenamefont {Seno},\ and\ \citenamefont
  {Maritan}}]{kocillari_signature_2018}%
  \BibitemOpen
  \bibfield  {author} {\bibinfo {author} {\bibfnamefont {L.}~\bibnamefont
  {Ko\c cillari}}, \bibinfo {author} {\bibfnamefont {P.}~\bibnamefont
  {Fariselli}}, \bibinfo {author} {\bibfnamefont {A.}~\bibnamefont {Trovato}},
  \bibinfo {author} {\bibfnamefont {F.}~\bibnamefont {Seno}},\ and\ \bibinfo
  {author} {\bibfnamefont {A.}~\bibnamefont {Maritan}},\ }\bibfield  {title}
  {\bibinfo {title} {Signature of {Pareto} optimization in the {Escherichia}
  coli proteome},\ }\href {https://doi.org/10.1038/s41598-018-27287-3}
  {\bibfield  {journal} {\bibinfo  {journal} {Sci Rep}\ }\textbf {\bibinfo
  {volume} {8}},\ \bibinfo {pages} {9141} (\bibinfo {year} {2018})}\BibitemShut
  {NoStop}%
\bibitem [{\citenamefont {Cobey}\ and\ \citenamefont
  {Hensley}(2017)}]{cobey_immune_2017}%
  \BibitemOpen
  \bibfield  {author} {\bibinfo {author} {\bibfnamefont {S.}~\bibnamefont
  {Cobey}}\ and\ \bibinfo {author} {\bibfnamefont {S.~E.}\ \bibnamefont
  {Hensley}},\ }\bibfield  {title} {\bibinfo {title} {Immune history and
  influenza virus susceptibility},\ }\href
  {https://doi.org/10.1016/j.coviro.2016.12.004} {\bibfield  {journal}
  {\bibinfo  {journal} {Current Opinion in Virology}\ }\textbf {\bibinfo
  {volume} {22}},\ \bibinfo {pages} {105} (\bibinfo {year} {2017})}\BibitemShut
  {NoStop}%
\bibitem [{\citenamefont {Hall}(1992)}]{hall_bootstrap_1992}%
  \BibitemOpen
  \bibfield  {author} {\bibinfo {author} {\bibfnamefont {P.}~\bibnamefont
  {Hall}},\ }\href {https://doi.org/10.1007/978-1-4612-4384-7} {\emph {\bibinfo
  {title} {The {Bootstrap} and {Edgeworth} {Expansion}}}},\ Springer {Series}
  in {Statistics}\ (\bibinfo  {publisher} {Springer New York},\ \bibinfo
  {address} {New York, NY},\ \bibinfo {year} {1992})\BibitemShut {NoStop}%
\bibitem [{\citenamefont {Blinnikov}\ and\ \citenamefont
  {Moessner}(1998)}]{blinnikov_expansions_1998}%
  \BibitemOpen
  \bibfield  {author} {\bibinfo {author} {\bibfnamefont {S.}~\bibnamefont
  {Blinnikov}}\ and\ \bibinfo {author} {\bibfnamefont {R.}~\bibnamefont
  {Moessner}},\ }\bibfield  {title} {\bibinfo {title} {Expansions for nearly
  {Gaussian} distributions},\ }\href {https://doi.org/10.1051/aas:1998221}
  {\bibfield  {journal} {\bibinfo  {journal} {Astron. Astrophys. Suppl. Ser.}\
  }\textbf {\bibinfo {volume} {130}},\ \bibinfo {pages} {193} (\bibinfo {year}
  {1998})}\BibitemShut {NoStop}%
\bibitem [{\citenamefont {Schweizer}(1992)}]{schweizer_meanvariance_1992}%
  \BibitemOpen
  \bibfield  {author} {\bibinfo {author} {\bibfnamefont {M.}~\bibnamefont
  {Schweizer}},\ }\bibfield  {title} {\bibinfo {title} {Mean-{Variance}
  {Hedging} for {General} {Claims}},\ }\bibfield  {journal} {\bibinfo
  {journal} {Ann. Appl. Probab.}\ }\textbf {\bibinfo {volume} {2}},\ \href
  {https://doi.org/10.1214/aoap/1177005776} {10.1214/aoap/1177005776} (\bibinfo
  {year} {1992})\BibitemShut {NoStop}%
\bibitem [{\citenamefont {Toft}(1996)}]{toft_meanvariance_1996}%
  \BibitemOpen
  \bibfield  {author} {\bibinfo {author} {\bibfnamefont {K.~B.}\ \bibnamefont
  {Toft}},\ }\bibfield  {title} {\bibinfo {title} {On the {Mean}-{Variance}
  {Tradeoff} in {Option} {Replication} with {Transactions} {Costs}},\ }\href
  {https://doi.org/10.2307/2331181} {\bibfield  {journal} {\bibinfo  {journal}
  {The Journal of Financial and Quantitative Analysis}\ }\textbf {\bibinfo
  {volume} {31}},\ \bibinfo {pages} {233} (\bibinfo {year} {1996})}\BibitemShut
  {NoStop}%
\bibitem [{\citenamefont {Ditzler}\ \emph {et~al.}(2015)\citenamefont
  {Ditzler}, \citenamefont {Roveri}, \citenamefont {Alippi},\ and\
  \citenamefont {Polikar}}]{ditzler_learning_2015}%
  \BibitemOpen
  \bibfield  {author} {\bibinfo {author} {\bibfnamefont {G.}~\bibnamefont
  {Ditzler}}, \bibinfo {author} {\bibfnamefont {M.}~\bibnamefont {Roveri}},
  \bibinfo {author} {\bibfnamefont {C.}~\bibnamefont {Alippi}},\ and\ \bibinfo
  {author} {\bibfnamefont {R.}~\bibnamefont {Polikar}},\ }\bibfield  {title}
  {\bibinfo {title} {Learning in {Nonstationary} {Environments}: {A}
  {Survey}},\ }\href {https://doi.org/10.1109/MCI.2015.2471196} {\bibfield
  {journal} {\bibinfo  {journal} {IEEE Comput. Intell. Mag.}\ }\textbf
  {\bibinfo {volume} {10}},\ \bibinfo {pages} {12} (\bibinfo {year}
  {2015})}\BibitemShut {NoStop}%
\bibitem [{\citenamefont {Lin}\ \emph {et~al.}(1999)\citenamefont {Lin},
  \citenamefont {Saito},\ and\ \citenamefont {Levine}}]{lin_edgeworth_1999}%
  \BibitemOpen
  \bibfield  {author} {\bibinfo {author} {\bibfnamefont {J.-J.}\ \bibnamefont
  {Lin}}, \bibinfo {author} {\bibfnamefont {N.}~\bibnamefont {Saito}},\ and\
  \bibinfo {author} {\bibfnamefont {R.~A.}\ \bibnamefont {Levine}},\ }\bibfield
   {title} {\bibinfo {title} {Edgeworth approximation of the
  {Kullback}-{Leibler} distance towards problems in image analysis},\ }\href
  {http://www.math. ucdavis.edu/saito} {\bibfield  {journal} {\bibinfo
  {journal} {University of California, Davis, Tech. Rep}\ } (\bibinfo {year}
  {1999})}\BibitemShut {NoStop}%
\bibitem [{\citenamefont {Inglada}\ and\ \citenamefont
  {Mercier}(2007)}]{inglada_new_2007}%
  \BibitemOpen
  \bibfield  {author} {\bibinfo {author} {\bibfnamefont {J.}~\bibnamefont
  {Inglada}}\ and\ \bibinfo {author} {\bibfnamefont {G.}~\bibnamefont
  {Mercier}},\ }\bibfield  {title} {\bibinfo {title} {A {New} {Statistical}
  {Similarity} {Measure} for {Change} {Detection} in {Multitemporal} {SAR}
  {Images} and {Its} {Extension} to {Multiscale} {Change} {Analysis}},\ }\href
  {https://doi.org/10.1109/TGRS.2007.893568} {\bibfield  {journal} {\bibinfo
  {journal} {IEEE Trans. Geosci. Remote Sensing}\ }\textbf {\bibinfo {volume}
  {45}},\ \bibinfo {pages} {1432} (\bibinfo {year} {2007})}\BibitemShut
  {NoStop}%
\end{thebibliography}%
\end{document}